\documentclass[prd,preprint,
superscriptaddress,tightenlines,nofootinbib, eqsecnum]{revtex4-2}

\usepackage{amsmath}
\usepackage{amsfonts}
\usepackage{amssymb}
\usepackage{bm}
\usepackage{hyperref}
\usepackage{mathrsfs}
\usepackage{graphicx}

\usepackage{empheq}

\usepackage{ulem}
\usepackage[usenames]{color}

\allowdisplaybreaks

\definecolor{darkgreen}{rgb}{0,0.5,0}

\hypersetup{
 bookmarks=true,         % show bookmarks bar?
 unicode=false,          % non-Latin characters in Acrobat???s bookmarks
 pdftoolbar=true,        % show Acrobat???s toolbar?
 pdfmenubar=true,        % show Acrobat???s menu?
 pdffitwindow=false,     % window fit to page when opened
 pdfstartview={FitH},    % fits the width of the page to the window
 pdftitle={My title},    % title
 pdfauthor={Author},     % author
 pdfsubject={Subject},   % subject of the document
 pdfcreator={Creator},   % creator of the document
 pdfproducer={Producer}, % producer of the document
 pdfkeywords={keyword1} {key2} {key3}, % list of keywords
 pdfnewwindow=true,      % links in new window
 colorlinks=true,       % false: boxed links; true: colored links
 linkcolor=red,          % color of internal links
 citecolor=cyan,        % color of links to bibliography
 filecolor=magenta,      % color of file links
 urlcolor=darkgreen,           % color of external links
 linktocpage=true
}

\newcommand{\FP}{\mathop{\mathrm{FP}}_{B=0}}
\newcommand{\FPprop}{\mathop{\mathrm{FP}}_{B=0}\Box^{-1}_\text{ret}}

\DeclareSymbolFontAlphabet{\mathrsfs}{rsfs}
\DeclareMathAlphabet{\mathcal}{OMS}{cmsy}{m}{n}

\newcommand\calO{{\mathcal{O}}}

\newcommand{\dd}{\mathrm{d}}

\newcommand{\dI}{\mathrm{I}}
\newcommand{\dJ}{\mathrm{J}}
\newcommand{\dM}{\mathrm{M}}
\newcommand{\dS}{\mathrm{S}}
\newcommand{\dW}{\mathrm{W}}
\newcommand{\dX}{\mathrm{X}}
\newcommand{\dY}{\mathrm{Y}}
\newcommand{\dZ}{\mathrm{Z}}

\allowdisplaybreaks
\DeclareSymbolFontAlphabet{\mathrsfs}{rsfs}
\DeclareMathAlphabet{\mathcal}{OMS}{cmsy}{m}{n}

\begin{document}
	
\title{The Quadrupole Moment of Compact Binaries \\to the Fourth post-Newtonian Order:\\ From Source to Canonical Moment}

\author{Luc \textsc{Blanchet}}\email{luc.blanchet@iap.fr}
\affiliation{$\mathcal{G}\mathbb{R}\varepsilon{\mathbb{C}}\mathcal{O}$, 
	Institut d'Astrophysique de Paris,\\ UMR 7095, CNRS, Sorbonne Universit{\'e},\\
	98\textsuperscript{bis} boulevard Arago, 75014 Paris, France}
\affiliation{Institut de Physique Th\'eorique, Universit\'e Paris-Saclay,\\
CEA, CNRS, 91191 Gif-sur-Yvette, France}

\author{Guillaume \textsc{Faye}}\email{faye@iap.fr}
\affiliation{$\mathcal{G}\mathbb{R}\varepsilon{\mathbb{C}}\mathcal{O}$, 
	Institut d'Astrophysique de Paris,\\ UMR 7095, CNRS, Sorbonne Universit{\'e},\\
	98\textsuperscript{bis} boulevard Arago, 75014 Paris, France}
	
\author{Fran\c{c}ois \textsc{Larrouturou}}\email{francois.larrouturou@desy.de}
\affiliation{Deutsches Elektronen-Synchrotron DESY, Notkestr. 85, 22607 Hamburg, Germany}
\affiliation{$\mathcal{G}\mathbb{R}\varepsilon{\mathbb{C}}\mathcal{O}$, 
	Institut d'Astrophysique de Paris,\\ UMR 7095, CNRS, Sorbonne Universit{\'e},\\
	98\textsuperscript{bis} boulevard Arago, 75014 Paris, France}

\date{\today}

\begin{abstract} 
As a crucial step towards the completion of the fourth post-Newtonian (4PN) gravitational-wave generation from compact binary systems, we obtain the expressions of the so-called ``canonical'' multipole moments of the source in terms of the ``source'' and ``gauge'' moments. The canonical moments describe the propagation of gravitational waves outside the source's near zone, while the source and gauge moments encode explicit information about the matter source. Those two descriptions, in terms of two sets of canonical moments or in terms of six sets of source and gauge moments, are isometric. We thus construct the non-linear diffeomorphism between them up to the third post-Minkowskian order, and we exhibit the concrete expression of the canonical mass-type quadrupole moment at the 4PN order. This computation is one of the last missing pieces for the determination of the gravitational-wave phasing of compact binary systems at 4PN order.
\end{abstract}


\maketitle

\section{Overview and result}
\label{sec:introduction}

Establishing accurate gravitational wave (GW) templates is crucial for modern astronomy, as those constitute critical material for the signal analysis of ground-based detectors and, in the future, airborne ones~\cite{advLIGO2015,advVIRGO2014,KAGRA2018}.
A major technique is the post-Newtonian (PN) approximation, which allows depicting the inspiralling phase of compact binaries, and constitutes the basis for effective phenomenological methods such as EOB (effective-one-body) or IMR (inspiral-merger-ringdown), able to describe the late inspiral and merger phases (see~\cite{Maggiore, BlanchetLR, BuonSathya15, Porto16}). 

Crucial to building PN waveforms is the knowledge of the mass-type quadrupole moment at a high accuracy level. Such quantity has been computed, in the case of non-spinning compact binaries, at the increasingly high 1PN~\cite{WagW76, BS89}, 2PN~\cite{BDI95, BDIWW95, WW96, LMRY19}, 3PN~\cite{BIJ02, BI04mult, BDEI04, BDEI05dr} and finally 4PN~\cite{MHLMFB20, DDR_source, DDR_radiatif} orders. Similarly, the mass octupole and current quadrupole have been computed up to 3PN order~\cite{FBI15, HFB_courant}. 

At the 4PN order, the mass quadrupole moment has been regularized by means of dimensional regularization, and both UV and IR divergences have been properly renormalized~\cite{DDR_source, DDR_radiatif}. An interesting feature of this 4PN accuracy is the non-locality in time which appears in the near-zone quantities, due to the conservative GW tail effect, \emph{i.e.} the backscattering of the radiation onto the static space-time curvature generated by the source.

Importantly, the previous computations concern the \textit{source-type} moments, either mass moments $\dI_L$ or current ones $\dJ_L$ (where $L=i_1\cdots i_\ell$ involves $\ell$ spatial indices). The source moments are directly connected to the matter distribution in the source, being known as explicit closed form integrals over the matter plus gravitation pseudo stress-energy tensor. However, when considering gravitational waves emitted by the source, it is more convenient to use a different set of multipole moments called canonical and denoted $\dM_L$ and $\dS_L$. The canonical moments $\dM_L$ and $\dS_L$ are directly associated with the two usual polarization states of GR, which are the two physical degrees of freedom in the GW propagation.

The goal of the present paper is to connect the canonical moments $\{\dM_L, \dS_L\}$ to the source moments $\{\dI_L, \dJ_L\}$ and to four additional gauge moments $\{\dW_L, \dX_L, \dY_L, \dZ_L\}$, which parametrize a linear gauge transformation performed in the external part of the source's near zone. We refer to~\cite{B96, B98mult, BFIS08} for discussions and details about the matching procedure we employ to link the near zone to the external zone.

It is important to emphasize that the two descriptions in terms of canonical moments $\{\dM_L, \dS_L\}$ or in terms of source and gauge moments $\{\dI_L, \dJ_L, \dW_L, \dX_L, \dY_L, \dZ_L\}$ are physically equivalent, \textit{i.e.} describe the same physical matter source if and only if the canonical moments are related in a precise way to the source and gauge moments up to arbitrary high orders~\cite{B96}. Hence there is a coordinate transformation linking the two descriptions of the same source, which is a non-linear deformation of the linear gauge transformation parametrized by the gauge moments $\{\dW_L, \dX_L, \dY_L, \dZ_L\}$.

In this paper, we complete a missing step towards the knowledge of the 4PN orbital phasing for compact binary inspiral by obtaining the relation between the canonical mass quadrupole moment $\dM_{ij}$ and the corresponding source one $\dI_{ij}$ at the 4PN order. Such relation was known previously at the leading 2.5PN order~\cite{B96} and at the next-to-leading 3.5PN order~\cite{BFIS08}. Up to the 3.5PN order, the correction terms in the canonical quadrupole are quadratic in the multipole moments; the cubic corrections start at 4PN, and are thus the aim of the present computation. 
Let us recapitulate the complete result for the canonical quadrupole moment up to the 4PN order\footnote{Hereafter, we denote with a capital letter $L$ a multi-index with $\ell$ indices, \emph{e.g.} $\dI_L = \dI_{i_1i_2\ldots i_\ell}$, with angular brackets the symmetric-trace-free (STF) projection, \emph{e.g.} $\dW_{a\langle i}\dI_{j\rangle a}\equiv \text{STF}_{ij} [\dW_{ai}\dI_{ja}]$; we systematically use STF harmonics~\cite{Sachs61, Pi64, Th80, BD86}; in the spirit of~\cite{HFB_courant}, we use the shorthands $\dJ_{i\vert L} \equiv \varepsilon_{ii_\ell a}\dJ_{aL-1}$ and $\dZ_{i\vert L} \equiv \varepsilon_{ii_\ell a}\dZ_{aL-1}$ for current type moments; the superscript $(n)$ denotes $n$ time derivatives; $(-)^n$ stands for $(-1)^n$ and $c$ is the speed of light and $G$ the gravitational constant.}
\begin{align}\label{eq:resMij}
\dM_{ij} &= 
\dI_{ij}
+ \frac{4G}{c^5}\bigg[\dW^{(2)}\dI_{ij}-\dW^{(1)}\dI^{(1)}_{ij}\bigg]\nonumber\\
&\
+ \frac{4G}{c^7}\Bigg\{
\frac{4}{7}\dW^{(1)}_{a\langle i}\dI^{(3)}_{j\rangle a}
+ \frac{6}{7}\dW_{a\langle i}\dI^{(4)}_{j\rangle a}
- \frac{1}{7}\dY^{(3)}_{a\langle i}\dI_{j\rangle a}
- \dY_{a\langle i}\dI^{(3)}_{j\rangle a}
- 2\dX\,\dI^{(3)}_{ij}\nonumber\\
& \qquad\qquad
- \frac{5}{21}\dW^{(4)}_{a}\dI_{ija}
+ \frac{1}{63}\dW^{(3)}_{a}\dI^{(1)}_{ija}
- \frac{25}{21}\dY^{(3)}_{a}\dI_{ija}
- \frac{22}{63}\dY^{(2)}_{a}\dI^{(1)}_{ija}
+ \frac{5}{63}\dY^{(1)}_{a}\dI^{(2)}_{ija}\nonumber\\
& \qquad\qquad
+ 2\dW^{(3)}\dW_{ij}
+ 2\dW^{(2)}\dW^{(1)}_{ij}
- \frac{4}{3}\dW_{\langle i}\dW^{(3)}_{j\rangle}
+ 2\dW^{(2)}\dY_{ij}\nonumber\\
& \qquad\qquad
- 4\dW_{\langle i}\dY^{(2)}_{j\rangle}
- \frac{1}{3}\dZ^{(3)}_{a\vert \langle i}\dI_{j\rangle a}
+ \dZ_{a\vert \langle i}\dI^{(3)}_{j\rangle a}
- \frac{4}{9}\dW^{(3)}_a\dJ_{\langle i\vert j\rangle a}
+ \frac{4}{9}\dY^{(2)}_a\dJ_{\langle i\vert j\rangle a}
- \frac{8}{9}\dY^{(1)}_a\dJ^{(1)}_{\langle i\vert j\rangle a}\Bigg\}\nonumber\\
&
+ \frac{G^2\dM}{c^8}\Bigg\{
\frac{106}{21}\dM\dW_{ij}^{(3)}
+ \frac{1084}{105}\dM\dY_{ij}^{(2)} 
\nonumber\\
&\qquad\qquad +
\frac{17642}{945}\dW^{(3)}\dI_{ij}
-\frac{7018}{315}\dW^{(2)}\dI_{ij}^{(1)}
+\frac{806}{45}\dW^{(1)}\dI_{ij}^{(2)}
-\frac{2098}{135}\dW\,\dI_{ij}^{(3)}\nonumber\\
&\qquad\qquad + 16\,\dI_{ij} \int_0^{+\infty}\!\!\dd\tau\,\ln\left(\frac{c\tau}{2r_0}\right) \dW^{(4)}(t-\tau)
- 16\,\dI_{ij}^{(1)} \int_0^{+\infty}\!\!\dd\tau\,\ln\left(\frac{c\tau}{2r_0}\right)\dW^{(3)}(t-\tau)\Bigg\} \nonumber\\
&+ \calO \left(\frac{1}{c^9}\right)\,.
\end{align}
Here $\dM$ is the constant (ADM) total mass, $\dW_{ij}$ and $\dY_{ij}$ for instance denote the quadrupoles associated with the series of gauge moments $\dW_L$ and $\dY_L$, $\dW$ and $\dW_i$ are the monopole and dipole of $\dW_L$, and so on. All those quantities are evaluated at time $t$ except when specified otherwise. In addition, the source is considered to be stationary in the remote past so that all time derivatives of the moments vanish before some instant $-\mathcal{T}$. This means in particular that the hereditary integrals are well defined. The expression~\eqref{eq:resMij} is valid in the center-of-mass frame, for which the mass dipole moment $\dI_i$ vanishes. The 4PN cubic terms are new with this paper, which extends the previously computed 3.5PN quadratic relation~\cite{B96, BFIS08, FMBI12}. Note that the analogous relations for the mass octupole and current quadrupole moments (currently known only at the leading 2.5PN order) do not receive 3PN corrections and can be found in~\cite{BFIS08}.

Interestingly, the relation between the canonical and source/gauge moments is not local, as clear from the 4PN tail integrals appearing in the last line of~\eqref{eq:resMij}. The associated scale $r_0$ is unphysical and should naturally disappear from any physical results, such as the 4PN flux. This will be a stringent test for our computation, which will however have to wait for the complete calculation of the 4PN \textit{radiative-type} moment, directly observable at future null infinity. Notice however that the tail terms in~\eqref{eq:resMij} will be zero in the case of circular orbits; see Eqs.~(5.7) in~\cite{FMBI12}.

Another worthy remark is that the subtleties arising from the use of an IR dimensional regularization scheme for the mass quadrupole have been completely tackled and solved in~\cite{DDR_source,DDR_radiatif}. We can thus safely perform the computation in three dimensions, using the standard Hadamard regularization scheme.

Finally, after the result~\eqref{eq:resMij}, only one last step remains before getting the 4PN mass quadrupole: the three-dimensional computation of cubic non-linear terms called ``tails-of-memory'', entering the relation between the radiative quadrupole moment and the canonical one at 4PN order. This is left to future work.

The plan of this paper is as follows. After reminders about the multipolar-post-Minkowskian (MPM) formalism in Sec.~\ref{sec:MPM}, we describe the general method for relating the canonical moments to the source and gauge moments up to any post-Minkowskian (PM) order in Sec.~\ref{sec:method} (extending earlier works in~\cite{B96, BFIS08}). Finally Sec.~\ref{sec:implementation} is devoted to the practical implementation that led to the result~\eqref{eq:resMij}, together with required formulas for retarded integrals of non-linear source terms. The essential, but technical, near-zone expansion of tail integrals is presented in details in App.~\ref{app:NZ_tail}. The verification of our main result~\eqref{eq:resMij} \textit{via} an alternative procedure is relegated in App.~\ref{app:method_2}.

\section{The Multipolar-post-Minkowskian expansion}
\label{sec:MPM}

The MPM formalism~\cite{BD86, B87, BD92, B98quad} obtains the general solution of the Einstein field equations outside a matter source in the form of a post-Minkowskian expansion, with each PM coefficient expanded as a formal multipolar series. The gothic metric deviation from the Minkowski metric, $h^{\mu\nu} \equiv \sqrt{-g}\,g^{\mu\nu}-\eta^{\mu\nu}$ where $g^{\mu\nu}$ is the inverse of the usual covariant metric, $\eta^{\mu\nu}$, that of the Minkowski metric, and $g\equiv\text{det}(g_{\mu\nu})$, obeys the Einstein's vacuum field equations in harmonic coordinates,
\begin{subequations}\label{eq:Einstein_eq}
	\begin{align}
	\Box h^{\mu\nu} &= \Lambda^{\mu\nu}[h,\partial h,\partial^2h] 
	\,,\\
	\partial_\nu h^{\mu\nu} &= 0\,.
	\end{align}
\end{subequations}
Here $\Box$ is the flat d'Alembertian operator and the gravitational source term $\Lambda^{\mu\nu}$ is at least quadratic in $h$ and its first and second partial derivatives. In this paper, we will reason to any PM order for the general method, but in practical computations, as we are interested in the 3PM interaction, we will only use the leading quadratic $\sim h\partial^2 h+ \partial h\partial h$ and sub-leading cubic $\sim h\partial h\partial h$ pieces in the source term.

\subsection{The generic MPM algorithm}

The ``generic'' non-linear MPM solution of the field equations is searched in the form of a non-linear expansion in the field perturbation, labeled by the gravitational constant $G$.
Formally, it may be represented to arbitrary high orders by the asymptotic series:
\begin{equation}\label{eq:hgen}
h_{\text{gen}}^{\mu\nu} = \sum_{n=1}^{+\infty} G^n h^{\mu\nu}_{\text{gen}\, n}
\,.
\end{equation}
The starting point is the most general solution of the linearized Einstein's vacuum equations in harmonic coordinates, $\Box h_{\text{gen}\,1}^{\mu\nu}=\partial_\nu h_{\text{gen}\,1}^{\mu\nu}=0$, which can be written in terms of six sets of symmetric-trace-free (STF) multipole moments $\{\dI_L,\dJ_L,\dW_L,\dX_L,\dY_L,\dZ_L\}$, dispatched between a simpler linear solution called ``canonical'', and a linear gauge transformation, as
\begin{equation}\label{eq:h1gen}
h_{\text{gen}\,1}^{\mu\nu}
= h_{\text{can}\,1}^{\mu\nu}
+ \partial\varphi_1^{\mu\nu}\,.
\end{equation}
We employ the shorthand notation $\partial\varphi_1^{\mu\nu} \equiv \partial^{\mu}\varphi_1^{\nu} + \partial^{\nu}\varphi_1^{\mu} - \eta^{\mu\nu}\,\partial_\rho\varphi_1^\rho$ for the linear gauge transformation. Denoting functionals of multipole moments by means of capital calligraphic letters, the functional dependence of the two terms in~\eqref{eq:h1gen} are
\begin{subequations}\begin{align}
	h_{\text{can}\,1}^{\mu\nu} &= \mathcal{H}_{\text{can}\,1}^{\mu\nu}\left[\dI_L,\dJ_L\right]\,,\\[0.2cm]
	\partial\varphi_1^{\mu\nu} &= \partial\Phi_1^{\mu\nu}\left[\dW_L,\dX_L,\dY_L,\dZ_L\right]\,.\label{eq:gauge1}
	\end{align}
\end{subequations}
For obvious reasons, the moments $\{\dI_L,\dJ_L\}$ (mass-type $I_L$ and current-type $J_L$) are called the source moments while the moments $\{\dW_L,\dX_L,\dY_L,\dZ_L\}$ are the gauge moments. The linear canonical solution, evaluated at field point $\mathbf{x}$ and at time $t$, reads explicitly~\cite{SB58, Pi64, Th80, BD86}
\begin{subequations}\label{eq:h1can}
	\begin{align}
	h_{\text{can}\,1}^{00} &= - \frac{4}{c^2}\sum_{\ell \geqslant 0}\frac{(-)^\ell}{\ell !}\partial_L\left[\frac{1}{r}\,\dI_L\left(t-\frac{r}{c}\right)\right]\,,\\
	%%%%%%%%%%%%%%%%%%%%%%%%%%%%%%%%%%%%%%%%%%%%%%%%%%%%%%%%%
	h_{\text{can}\,1}^{0i} &= \frac{4}{c^3}\sum_{\ell \geqslant 1}\frac{(-)^\ell}{\ell !}\left\lbrace\partial_{L-1}\left[\frac{1}{r}\,\dI_{iL-1}^{(1)}\left(t-\frac{r}{c}\right)\right]+ \frac{\ell}{\ell+1}\partial_L\left[\frac{1}{r}\,\dJ_{i\vert L}\left(t-\frac{r}{c}\right)\right]\right\rbrace\,,\\
	%%%%%%%%%%%%%%%%%%%%%%%%%%%%%%%%%%%%%%%%%%%%%%%%%%%%%%%%%
	h_{\text{can}\,1}^{ij} &= -\frac{4}{c^4}\sum_{\ell \geqslant 2}\frac{(-)^\ell}{\ell !}\left\lbrace\partial_{L-2}\left[\frac{1}{r}\,\dI_{ijL-2}^{(2)}\left(t-\frac{r}{c}\right)\right]+ \frac{2\ell}{\ell+1}\partial_{L-1}\left[\frac{1}{r}\,\dJ_{(i\vert j) L-1}^{(1)}\left(t-\frac{r}{c}\right)\right]\right\rbrace\,,
	\end{align}
\end{subequations}
where $r=|\mathbf{x}|$ represents the radial distance to the origin located in the source. From the harmonic coordinate condition $\partial_\nu h_{\text{can}\,1}^{\mu\nu}=0$, the mass monopole $I$ and current dipole $J_i$ must be constant, while the mass dipole $I_i$ is varying linearly with time. In applications, we choose a center-of-mass frame for which $I_i=0$.\footnote{Note that the mass dipole $I_i$ is defined here in an ADM sense and, as such, includes both matter and gravitational contributions. Thus the displacement of the center-of-mass due to gravitational radiation, which concerns the matter system and begins at the 3.5PN order (see \textit{e.g.}~\cite{BF18}), is already included in $I_i$ and is compensated by the contribution of radiation.} With these exceptions, the moments are arbitrary functions of time encoding the properties of the source. The linearized gauge vector is defined by
\begin{subequations}\label{eq:phi1}
	\begin{align} 
	\varphi_1^0 =&\
	\frac{4}{c^3}\sum_{\ell \geqslant 0}\frac{(-)^\ell}{\ell !}\partial_L\left[\frac{1}{r}\,\dW_L\left(t-\frac{r}{c}\right)\right]\,,\\
	\varphi_1^i =&
	-\frac{4}{c^4}\sum_{\ell \geqslant 0}\frac{(-)^\ell}{\ell !}\partial_{iL}\left[\frac{1}{r}\,\dX_L\left(t-\frac{r}{c}\right)\right]\nonumber\\
	&
	-\frac{4}{c^4}\sum_{\ell \geqslant 1}\frac{(-)^\ell}{\ell !}\left\lbrace\partial_{L-1}\left[\frac{1}{r}\,\dY_{iL-1}\left(t-\frac{r}{c}\right)\right]+ \frac{\ell}{\ell+1}\partial_L\left[\frac{1}{r}\,\dZ_{i\vert L}\left(t-\frac{r}{c}\right)\right]\right\rbrace\,.
	\end{align}
\end{subequations}
The gauge moments $\{\dW_L,\dX_L,\dY_L,\dZ_L\}$ are arbitrary functions of time without restriction.

The MPM construction is defined by induction on the PM order $n$~\cite{BD86}. Suppose that for some given $n\geqslant 2$, one has obtained the first $n-1$ PM coefficients $h_{\text{gen}\,m}^{\mu\nu}$, $\forall m\leqslant n-1$. Then, the next order coefficient $h_{\text{gen}\,n}^{\mu\nu}$ is constructed as follows. It satisfies  
\begin{subequations}\label{eq:Boxhn}
	\begin{align}
	\Box h_{\text{gen}\,n}^{\mu\nu} &= \Lambda_{\text{gen}\,n}^{\mu\nu} \equiv \Lambda^{\mu\nu}_n \left[h_{\text{gen}\,1}, \cdots, h_{\text{gen}\,n-1}\right]\,,\\[0.2cm]
	\partial_\nu h_{\text{gen}\,n}^{\mu\nu} &= 0\,,
	\end{align}
\end{subequations}
where the source term $\Lambda_{\text{gen}\,n}^{\mu\nu}$, being at least quadratic in $h$, depends only on the previous iterations as indicated. We first construct a particular retarded solution of the wave equation $\Box h_{\text{gen}\,n}^{\mu\nu} = \Lambda_{\text{gen}\,n}^{\mu\nu}$ as
\begin{equation}\label{eq:ungen}
u_{\text{gen}\, n}^{\mu\nu} 
\equiv 
\FPprop\left[\left(\frac{r}{r_0}\right)^B\Lambda_{\text{gen}\, n}^{\mu\nu} \right]\,,
\end{equation}
where $\Box^{-1}_\text{ret}$ denotes the usual retarded inverse d'Alembertian operator, and the symbol $\FP$ refers to a specific operation of taking the finite part (FP) in the Laurent expansion when the complex parameter $B$ tends to zero. This finite part involves the multiplication of the source term by the regularization factor $(r/r_0)^B$, where we introduce an arbitrary constant length scale $r_0$. Such FP operation is required for dealing with source terms made with multipolar expansions like in~\eqref{eq:h1can} that are singular at the origin $r = 0$. More generally, the regularized retarded integral operator $\mathrm{FP}\, \Box^{-1}_\mathrm{ret}\equiv\FPprop (r/r_0)^B$ is well defined when acting on a source term admitting an expansion when $r\to 0$ of the form, for any $N\in\mathbb{N}$,
\begin{align}\label{eq:structF}
f(\mathbf{x},t) = \sum_{\ell=0}^{+\infty} \,\sum_{a=a_\text{min}}^{N}\,\sum_{p=0}^{p_\text{max}} f^L_a(t) \,\hat{n}_L \,r^a \ln^p r + o\left(r^N\right)\, ,
\end{align}
where $\hat{n}_L= \text{STF} [n_{i_1} n_{i_2}\cdots n_{i_\ell}]$, with $n_i=x^i/r$, is the STF spherical harmonics of order $\ell$, while the sum boundaries $a_\text{min}$ and $p_\text{max}$ are integers (depending on the PM order $n$). For any function in the class~\eqref{eq:structF}, we have $\Box[\mathrm{FP}\, \Box^{-1}_\mathrm{ret} f]=f$. In the end of the recursive process, the structure~\eqref{eq:structF} turns out to be proved by induction.

Because of the regularization scheme, the object $u_{\text{gen}\, n}^{\mu\nu}$ does not satisfy the harmonic gauge condition, but a simple calculation using the fact that the source term is divergenceless, $\partial_\nu \Lambda_{\text{gen}\, n}^{\mu\nu} = 0$, gives
\begin{equation}\label{eq:wngen}
w_{\text{gen}\, n}^{\mu} 
\equiv \partial_\nu u_{\text{gen}\, n}^{\mu\nu} = 
\FPprop\left[B\left(\frac{r}{r_0}\right)^B\frac{n_i}{r}\,\Lambda_{\text{gen}\, n}^{\mu i} \right]\,.
\end{equation}
Due to the explicit factor $B$, this term is non zero only when the integral generates a pole $\propto 1/B$ in the Laurent expansion as $B\to 0$. In turn, the pole arises only from the singular behaviour of the source term as $r\to 0$, which is of the type~\eqref{eq:structF}. Furthermore, the coefficient of the pole is necessarily a homogeneous retarded solution of the wave equation, since the source term has no pole, hence $\Box w_{\text{gen}\, n}^{\mu} = 0$. At this stage, we apply the MPM ``harmonicity'' algorithm to construct from $w_{\text{gen}\, n}^{\mu}$ another homogeneous retarded solution, say 
\begin{equation}\label{eq:vngen}
v_{\text{gen}\, n}^{\mu\nu} \equiv \mathcal{V}^{\mu\nu}\bigl[w_{\text{gen}\, n}\bigr]\,,
\end{equation}
satisfying $\partial_\mu v_{\text{gen}\, n}^{\mu\nu} = - w_{\text{gen}\, n}^{\mu}$, together with $\Box v_{\text{gen}\, n}^{\mu\nu} = 0$. The formulas specifying the above harmonicity algorithm $w^\mu\longrightarrow\mathcal{V}^{\mu\nu}[w]$ are given by \textit{e.g.} Eqs.~(2.11)-(2.12) in~\cite{B98quad}. Finally, the metric at the PM order $n$, now satisfying the full Einstein vacuum equations in harmonic coordinates at the PM order $n$, is naturally defined as
\begin{equation}\label{eq:hngen}
h_{\text{gen}\, n}^{\mu\nu} = u_{\text{gen}\, n}^{\mu\nu} + v_{\text{gen}\, n}^{\mu\nu}\,.
\end{equation}

This construction yields the most general solution $h_{\text{gen}}^{\mu\nu}$ of the Einstein field equations in the vacuum region external to any isolated matter system~\cite{BD86}. It is obtained in the form of a functional of six sets of multipole moments,
\begin{equation}\label{eq:hgenstruct}
h^{\mu\nu}_{\text{gen}} = \sum_{n=1}^{+\infty} G^n h^{\mu\nu}_{\text{gen}\,n} = \mathcal{H}^{\mu\nu}_{\text{gen}}\left[\dI_L, \dJ_L, \dW_L, \dX_L, \dY_L, \dZ_L\right]
\,.
\end{equation} 

\subsection{The canonical MPM algorithm}

However, we also know that the general field of an isolated system in GR can be described by two and only two sets of STF multipole moments, which we call the ``canonical'' moments and denote $\{\dM_L, \dS_L\}$. Notably, these moments describe the GW propagation far away from the source, and they parametrize the two GW tensorial modes of GR~\cite{BD86}. The description of the external field of the source in terms of $\{\dM_L, \dS_L\}$ starts at linear order just by the canonical metric~\eqref{eq:h1can} instead of the generic metric~\eqref{eq:h1gen}, but, in this different set up, the linear approximation is parametrized by the canonical moments $\dM_L$ and $\dS_L$:
\begin{equation}\label{eq:h1canMS}
h_{\text{can}\,1}^{\mu\nu}
= \mathcal{H}_{\text{can}\,1}^{\mu\nu}\left[\dM_L,\dS_L\right]\,.
\end{equation}
The canonical MPM algorithm proceeds then by induction over the PM order $n\geqslant 2$ exactly in the same way as before, \textit{i.e.} following the synthetic steps
\begin{align}\label{eq:hncan}
&\left.\begin{array}{l} \displaystyle u_{\text{can}\, n}^{\mu\nu} 
=
\FPprop\biggl[\left(\frac{r}{r_0}\right)^B\Lambda_{\text{can}\, n}^{\mu\nu} \biggr] \\[0.2cm]\displaystyle w_{\text{can}\, n}^{\mu} = \partial_\nu u_{\text{can}\, n}^{\mu\nu} = 
\FPprop\biggl[B\left(\frac{r}{r_0}\right)^B\frac{n_i}{r}\,\Lambda_{\text{can}\, n}^{\mu i} \biggr] \\[0.4cm]v_{\text{can}\, n}^{\mu\nu} = \mathcal{V}^{\mu\nu}\bigl[w_{\text{can}\, n}\bigr]
\end{array}\right\} \Longrightarrow ~h_{\text{can}\, n}^{\mu\nu} = u_{\text{can}\, n}^{\mu\nu} + v_{\text{can}\, n}^{\mu\nu}\,.
\end{align}
This results in a full non-linear metric, which is a functional of the canonical moments only, and represents as well the most general solution of the Einstein field equations outside the source:
\begin{align}\label{eq:hcan}
h^{\mu\nu}_{\text{can}} &= \sum_{n=1}^{+\infty} G^n h^{\mu\nu}_{\text{can}\,n} = \mathcal{H}^{\mu\nu}_{\text{can}}\left[\dM_L, \dS_L\right]\,.
\end{align}
The next section presents the general method to obtain the relationships linking the canonical moments $\{\dM_L, \dS_L\}$ to the set of source and gauge moments $\{\dI_L, \dJ_L, \cdots, \dZ_L\}$, valid in principle to any PM order.
We thus apply this general method to the practical case of cubic interaction to derive the 4PN relation~\eqref{eq:resMij}.

\section{Relation between canonical and source/gauge moments}
\label{sec:method}

\subsection{General method}\label{sec:general_method}

In Sec.~\ref{sec:MPM}, we constructed two full non-linear MPM solutions, Eqs.~\eqref{eq:hgen} and~\eqref{eq:hcan}, which both represent the most general solution of the Einstein field equations in the vacuum region outside the source. Requiring that these two metrics describe the exterior field of the \emph{same} physical system, we now impose that they are physically equivalent, \textit{i.e.}, just differ by a coordinate transformation. This implies unique relations between the canonical moments $\{\dM_L, \dS_L\}$ and the source and gauge moments $\{\dI_L, \dJ_L, \dW_L, \dX_L, \dY_L, \dZ_L\}$, which can be viewed as two physically equivalent sets of moments for describing the source as seen from its exterior.

We thus look for a coordinate transformation $x^\mu\longrightarrow
x'^\mu$ such that
\begin{equation}\label{eq:coordtransf}
h^{\mu\nu}_\text{gen}(x') = \frac{1}{\vert J\vert} \,\frac{\partial
x'^\mu}{\partial x^\rho}\,\frac{\partial x'^\nu}{\partial
x^\sigma}\Bigl(h^{\rho\sigma}_\text{can}(x) + \eta^{\rho\sigma}\Bigr) -
\eta^{\mu\nu}  \,,
\end{equation}
where $J\equiv\text{det}(\partial x'/\partial x)$ is the Jacobian of the transformation. Eq.~\eqref{eq:coordtransf} immediately follows from the definition $h^{\mu\nu} = \sqrt{-g}\,g^{\mu\nu}-\eta^{\mu\nu}$ and the law of transformation of tensors~\cite{MTW}. Introducing a coordinate shift $\varphi^\mu$ such that $x'^\mu = x^\mu + \varphi^\mu(x)$, we can rewrite the statement~\eqref{eq:coordtransf}, using the non-linear correction $\delta_{\varphi}h^{\mu\nu}_\text{can}(x)$ to the metric $h^{\mu\nu}_\text{can}(x)$ induced by the shift, as
\begin{equation}\label{eq:Lie}
h^{\mu\nu}_\text{gen}(x) = h^{\mu\nu}_{\text{can}}(x) + \delta_\varphi h^{\mu\nu}_{\text{can}}(x)\,.
\end{equation}
It is implicit that we work perturbatively in both the vector $\varphi^\lambda$ and the metric $h^{\lambda\rho}_{\text{can}}$, so that we have for instance $h^{\mu\nu}_\text{gen}(x') = \sum_{n\geqslant 0} \varphi^{\lambda_1}\cdots\varphi^{\lambda_n}\partial_{\lambda_1}\cdots\partial_{\lambda_n}h^{\mu\nu}_\text{gen}(x)/n!$. To linear order, the correction reduces to $\partial\varphi^{\mu\nu} = \partial^{\mu}\varphi^{\nu} + \partial^{\nu}\varphi^{\mu} - \eta^{\mu\nu}\,\partial_\rho\varphi^\rho$. Let us then pose, to any order,
\begin{equation}\label{eq:Liedetail}
\delta_\varphi h^{\mu\nu}_{\text{can}} = \partial\varphi^{\mu\nu} + \Omega^{\mu\nu}\left[\varphi, h_{\text{can}}\right]\,,
\end{equation}
where $\Omega^{\mu\nu}$ denotes a functional of $\varphi^\lambda$ and $h^{\lambda\rho}_{\text{can}}$, as well as their derivatives, which is at least quadratic and can be computed perturbatively up to any order using Eq.~\eqref{eq:coordtransf}. The harmonic gauge condition satisfied by both metrics $h^{\mu\nu}_{\text{gen}}$ and $h^{\mu\nu}_{\text{can}}$ implies that (as a consequence of the identity $\partial_\nu\partial\varphi^{\mu\nu}=\Box\varphi^{\mu}$)
\begin{align}\label{eq:harmcoord} 
\Box \varphi^\mu +
\partial_\nu \Omega^{\mu\nu} =0\, .
\end{align}

Let us now look for the coordinate shift in the form of a full PM expansion series
\begin{equation}\label{eq:varphimu}
\varphi^\mu = \sum_{n=1}^{+\infty} G^n \varphi^\mu_n\,,
\end{equation}
and, conjointly, for the relations between the canonical moments and the source/gauge moments in the same PM form
	\begin{equation}\label{eq:ML_SL}
	\dM_L = \sum_{n=1}^{+\infty} G^{n-1}
	\dM_{n\,L}\,,\qquad \dS_L = \sum_{n=1}^{+\infty} G^{n-1} \dS_{n\,L}\,.
	\end{equation}
Here $\dM_{n\,L}$ and $\dS_{n\,L}$ denote some $n$-th non-linear functionals of the six types of source and gauge moments $\dI_K$, $\cdots$, $\dZ_K$ (with, say, $K=i_1\cdots i_k$), which are to be determined: 
\begin{equation}\label{eq:functML_SL}
\dM_{n\,L} = \mathcal{M}_{n\,L}\left[\dI_K, \dJ_K, \cdots, \dZ_K\right]\,,\qquad\dS_{n\,L} = \mathcal{S}_{n\,L}\left[\dI_K, \dJ_K, \cdots, \dZ_K\right]\,.
\end{equation}
At linear order, as we have seen with Eq.~\eqref{eq:h1gen}, by definition of the two linear approximations for the two generic and canonical metrics,
\begin{equation}\label{eq:h1genv2}
h_{\text{gen}\,1}^{\mu\nu}
= \mathcal{H}_{\text{can}\,1}^{\mu\nu}\left[\dI_L, \dJ_L\right] + \partial\varphi_1^{\mu\nu}\,,
\end{equation}
where the functional $\mathcal{H}_{\text{can}\,1}^{\mu\nu}$ is explicitly given by~\eqref{eq:h1can} and the gauge vector is parametrized by the gauge moments $\{\dW_L, \cdots, \dZ_L\}$ [see its expression~\eqref{eq:phi1}]. Eq.~\eqref{eq:h1genv2} means that the relation~\eqref{eq:Lie} is satisfied at leading order provided that (i) $\varphi_{1}^{\mu}= \Phi_{1}^{\mu}[\dW_L, \dX_L, \dY_L, \dZ_L]$ and (ii) the moments $\{\dM_L, \dS_L\}$ reduce to the source moments $\{\dI_L, \dJ_L\}$ to leading order, \textit{i.e.}
\begin{align}
\dM_{1\,L} = \dI_L\,,\qquad \dS_{1\,L} = \dJ_L \,.
\end{align}
Thus, we see that the looked-for coordinate transformation will be a non-linear deformation of the linear gauge transformation associated with the shift~\eqref{eq:phi1}.  

Our recurrence hypothesis will be that all the $\varphi^\mu_m$'s in~\eqref{eq:varphimu} are known up to a given PM order $n-1$, as are the functional relations~\eqref{eq:ML_SL} up to the corresponding order. Hence we assume that we have already determined
\begin{subequations}\label{eq:ML_SLn}
	\begin{align}
	\varphi^\mu_{\leqslant n-1} &\equiv \sum_{m=1}^{n-1} G^m \varphi^\mu_m\,,\\
	\dM_{\leqslant n-1\,L} &\equiv \sum_{m=1}^{n-1} G^{m-1}
	\mathcal{M}_{m\,L}\left[\dI_K, \dJ_K, \cdots, \dZ_K\right]\,,\\ \dS_{\leqslant n-1\,L} &\equiv \sum_{m=1}^{n-1} G^{m-1} \mathcal{S}_{m\,L}\left[\dI_K, \dJ_K, \cdots, \dZ_K\right]\,,
	\end{align}
\end{subequations}
in such a way that the equations~\eqref{eq:Lie}--\eqref{eq:Liedetail} hold up to order $n-1$. This means that, for all $m\leqslant n-1$,
\begin{equation}
h^{\mu\nu}_{\text{gen}\,m} = h^{\mu\nu}_{\text{can}\,m} + \partial\varphi^{\mu\nu}_m + \Omega^{\mu\nu}_m\bigl[\varphi_1\cdots\varphi_{m-1}; h_{\text{can}\,1}\cdots h_{\text{can}\,m-1}\bigr]\,.
\end{equation}
Recall that $\Omega^{\mu\nu}_m$ is a non-linear, at least quadratic, functional of the coordinate shift and metric, and is therefore known following our induction hypothesis.
Indeed, using \eqref{eq:Liedetail} and~\eqref{eq:h1genv2}, we get $\Omega^{\mu\nu}_1\equiv 0$. Crucial in our hypothesis is the assumption that the canonical metric depends on the moments so far determined to order $n-1$:
\begin{align}\label{eq:Hcanm}
h^{\mu\nu}_{\text{can}\,m} \equiv \mathcal{H}^{\mu\nu}_{\text{can}\,m}\left[\dM_{\leqslant n-1\,L}, \dS_{\leqslant n-1\,L}\right]\,.
	\end{align}

\subsection{Implementation to \boldmath $n$\unboldmath PM order}
\label{sec:implnPM}

With our recurrence hypothesis, let us see how to the next PM order $\varphi^{\mu}_{n}$ together with the functionals $\mathcal{M}_{n\, L}$ and $\mathcal{S}_{n\, L}$ are uniquely determined. We thus want to find $\varphi^{\mu}_{n}$ and $\mathcal{M}_{n\, L}$, $\mathcal{S}_{n\, L}$ such that
\begin{equation}\label{eq:ordren}
h^{\mu\nu}_{\text{gen}\,n} = h^{\mu\nu}_{\text{can}\,n} + \partial\varphi^{\mu\nu}_n + \Omega^{\mu\nu}_n\,,
\end{equation}
where $h^{\mu\nu}_{\text{gen}\,n}$ and $h^{\mu\nu}_{\text{can}\,n}$ are defined precisely by the two MPM constructions in Sec.~\ref{sec:MPM}, and $h^{\mu\nu}_{\text{can}\,n}$ is now a functional of $\dM_{\leqslant n\,L}$ and $\dS_{\leqslant n\,L}$. First of all, note that $\Omega^{\mu\nu}_n$ depends on $\varphi_k$ and $h_{\text{can}\,k}$ for $k\leqslant n-1$ and is already known by our induction hypothesis. Second, apply the harmonic coordinate conditions on Eq.~\eqref{eq:ordren}: this shows that while $\varphi^{\mu}_{n}$ is one of our unknowns, its d'Alembertian $\Delta^{\mu}_{n}\equiv\Box\varphi^{\mu}_{n}$ is already determined as we have [see Eq.~\eqref{eq:harmcoord}]
\begin{equation}\label{eq:divordren}
\Delta^{\mu}_{n} + \partial_\nu\Omega^{\mu\nu}_n = 0\,.
\end{equation}
The explicit expressions of $\Omega^{\mu\nu}_n$ and $\Delta^{\mu}_{n}$ at quadratic and cubic orders are displayed in Eqs.~\eqref{eq:OmegaDelta},~\eqref{eq:Omega3} and~\eqref{eq:Delta3}.
At this stage, we must use the specific definitions of the generic and canonical metrics defined in Sec.~\ref{sec:MPM}. Applying the d'Alembertian operator on~\eqref{eq:ordren}, we find that the two source terms $\Lambda^{\mu\nu}_{\text{gen}\,n}$ and $\Lambda^{\mu\nu}_{\text{can}\,n}$ are related by
\begin{equation}\label{eq:sourcen}
\Lambda^{\mu\nu}_{\text{gen}\,n} = \Lambda^{\mu\nu}_{\text{can}\,n} + \partial\Delta^{\mu\nu}_n + \Box\Omega^{\mu\nu}_n\,.
\end{equation}
Hence, the particular retarded solutions of the two algorithms, $u^{\mu\nu}_{\text{gen}\,n}$ and $u^{\mu\nu}_{\text{can}\,n}$, defined respectively in~\eqref{eq:ungen} and~\eqref{eq:hncan}, satisfy
\begin{equation}\label{eq:ungencan0}
u^{\mu\nu}_{\text{gen}\,n} = u^{\mu\nu}_{\text{can}\,n} + \FPprop \biggl[
\left(\frac{r}{r_0}\right)^B\Bigr(\partial\Delta^{\mu\nu}_n + \Box\Omega^{\mu\nu}_n\Bigr)\biggr]\,.
\end{equation}
Next, we introduce a linear-looking gauge transformation with vector defined by the retarded integral of $\Delta_n^{\mu}$, as
\begin{equation}\label{eq:phin}
\phi_n^{\mu} = \FPprop \biggl[
\left(\frac{r}{r_0}\right)^B \Delta_n^{\mu} \biggr]\,.
\end{equation}
This vector satisfies $\Box\phi_n^{\mu} = \Delta_n^{\mu}$ but is not yet our looked-for vector $\varphi_n^{\mu}$. It a priori differs from it by an homogeneous retarded solution of the wave equation. Thanks to~\eqref{eq:phin} we can advantageously rewrite~\eqref{eq:ungencan0} as
\begin{equation}\label{eq:ugencann}
u_{\mathrm{gen}\,n}^{\mu\nu} = u_{\mathrm{can}\,n}^{\mu\nu} + \partial\phi_n^{\mu\nu} +
\Omega_n^{\mu\nu} + X_n^{\mu\nu} + Y_n^{\mu\nu} \,.
\end{equation}
The last two terms are the most interesting: they come from the non-commutation of the finite part of the retarded integral $\FPprop$ with the partial derivative, due to the differentiation of the regularization factor $(r/r_0)^B$ therein. They are thus given by
\begin{subequations}\label{eq:commXYdef}
	\begin{align}
	X_n^{\mu\nu} &\equiv \FPprop \left[ \left(\frac{r}{r_0}\right)^B \Box \Omega_n^{\mu\nu}\right] -
	\Omega_n^{\mu\nu} \,,\\ 
	Y_n^{\mu\nu} &\equiv \FPprop \left[ \left(\frac{r}{r_0}\right)^B
	\partial\Delta_n^{\mu\nu}\right] - \partial\phi_n^{\mu\nu}  \,.
	\end{align}
\end{subequations}
Observing that, for the class of multipole-expanded functions $f$ we are concerned with [see Eq.~\eqref{eq:structF}], the statement $\Box(\mathrm{FP}\,\Box^{-1}_\mathrm{ret}f)=f$ is always correct, we see that $X_n^{\mu\nu}$ and $Y_n^{\mu\nu}$ represent the ``commutators'' of the operators that appear inside the square brackets:
\begin{subequations}\label{eq:commXYdef2}
	\begin{align}
	X_n^{\mu\nu} &= \Bigl[\mathrm{FP}\, \Box^{-1}_\mathrm{ret}, \,\Box\Bigr]
	\,\Omega_n^{\mu\nu}\,,\\ 
	Y_n^{\mu\nu} &= \Bigl[\mathrm{FP}\,
	\Box^{-1}_\mathrm{ret}, \,\partial\Bigr]\,\Delta_n^{\mu\nu} \,.
	\end{align}
\end{subequations}
Since the differentiation of the regularization factor $(r/r_0)^B$ produces an extra factor $B$, the quantities $X_n^{\mu\nu}$ and $Y_n^{\mu\nu}$ will be non-zero only when the integral develops a pole $\sim 1/B$. In that case, they are necessarily homogeneous (retarded) solutions of the wave equation: $\Box X_n^{\mu\nu} = \Box Y_n^{\mu\nu} =0$. It is easy to figure out that if $X_n^{\mu\nu}$ and $Y_n^{\mu\nu}$ were actually zero, the canonical moments $\{\dM_L, \dS_L\}$ would simply agree with their source counterparts $\{\dI_L, \dJ_L\}$. As a result, the non trivial relations between those moments entirely follow from the evaluation of the two quantities $X_n^{\mu\nu}$ and $Y_n^{\mu\nu}$. In our practical calculations, we reshuffle the commutators~\eqref{eq:commXYdef2} and use the following expressions, exhibiting the explicit factor $B$ in front:
\begin{subequations}\label{eq:commXY}\begin{align}
	X_n^{\mu\nu} &= \FPprop \biggl[ B \left(\frac{r}{r_0}\right)^B
	\biggl(-\frac{B+1}{r^{2}}\,\Omega_n^{\mu\nu} -
	\frac{2}{r}\,\partial_r \Omega_n^{\mu\nu}\biggr)\biggr]\,,\\ 
	Y_n^{\mu\nu} &= \FPprop \biggl[ B \left(\frac{r}{r_0}\right)^B \frac{n_i}{r}
	\biggl( - \delta^{i\mu} \Delta_n^{\nu} - \delta^{i\nu} \Delta_n^{\mu} +
	\eta^{\mu\nu}\Delta_n^i\biggr) \biggr]\,.
	\end{align}\end{subequations}
Next, we carry on the MPM algorithm by computing the divergence of~\eqref{eq:ugencann}. With Eq.~\eqref{eq:divordren}, we obtain $w_{\mathrm{gen}\,n}^{\mu} = w_{\mathrm{can}\,n}^{\mu} + W_n^{\mu}$, where we have posed $U_{n}^{\mu\nu} \equiv X_n^{\mu\nu} + Y_n^{\mu\nu}$ and $W_{n}^{\mu} \equiv \partial_\nu U_{n}^{\mu\nu}$. It is in fact necessary and sufficient to apply the harmonicity algorithm $\mathcal{V}^{\mu\nu}$ only to the divergence of the sum of the two commutators~\eqref{eq:commXY}. We have
\begin{align}\label{eq:hgenericn}
\left.\begin{array}{l} \displaystyle U_{n}^{\mu\nu} = X_n^{\mu\nu} + Y_n^{\mu\nu} \\[0.2cm]\displaystyle W_{n}^{\mu} = \partial_\nu U_{n}^{\mu\nu} \\[0.2cm]V_{n}^{\mu\nu} = \mathcal{V}^{\mu\nu}\bigl[W_{n}\bigr]
\end{array}\right\} \Longrightarrow ~h_{\mathrm{gen}\,n}^{\mu\nu} = h_{\mathrm{can}\,n}^{\mu\nu} + \partial\phi_n^{\mu\nu} + \Omega_n^{\mu\nu} + U_n^{\mu\nu} + V_n^{\mu\nu}\,.
\end{align}
The final step consists in remarking that $H_n^{\mu\nu}\equiv U_n^{\mu\nu} + V_n^{\mu\nu}$ is not only divergenceless by definition of the harmonicity algorithm $\mathcal{V}^{\mu\nu}$, but that it is also a retarded homogeneous solution of the wave equation, since both $U_n^{\mu\nu}$ and $V_n^{\mu\nu}$ are separately such. Hence $H_n^{\mu\nu}$ satisfies the linearized vacuum Einstein field equations, \textit{i.e.} $\Box H_n^{\mu\nu} = \partial_\nu H_n^{\mu\nu} = 0$, to which we know the general solution. Namely, it can be decomposed in a unique way as
\begin{equation}\label{eq:Hn}
H_n^{\mu\nu} = \mathcal{H}_{\mathrm{can}\,1}^{\mu\nu}\bigl[\dM_{n\,L}, \dS_{n\,L}\bigr] + \partial\psi_n^{\mu\nu}\,,
\end{equation}
where $\mathcal{H}_{\mathrm{can}\,1}^{\mu\nu}$ denotes the linearized functional~\eqref{eq:h1can} of the moments, but computed with certain moments $\dM_{n\,L}$ and $\dS_{n\,L}$, and where $\partial\psi_n^{\mu\nu}$ is some linear-looking gauge transformation. The associated gauge ``vector'' $\psi_n^{\mu}$ is parametrized in a unique way by some moments $\{\dW_{n\,L}, \dX_{n\,L}, \dY_{n\,L}, \dZ_{n\,L}\}$. The vector $\psi_n^{\mu}$ represents the homogeneous solution to be added to $\phi_n^{\mu}$ as given by Eq.~\eqref{eq:phin} in order to recover Eq.~\eqref{eq:ordren} with the shift 
\begin{equation}\label{eq:varphin}
\varphi_n^{\mu} \equiv \phi_n^{\mu} + \psi_n^{\mu}\,.
\end{equation}
On the other hand, it is clear that $\{\dM_{n\,L}, \dS_{n\,L}\}$ represent the looked-for corrections to the moments to the order $n$. Indeed, by the linearity of the functional $\mathcal{H}_{\mathrm{can}\,1}^{\mu\nu}$, the first term in~\eqref{eq:Hn} nicely combines with the linearized approximation in our induction hypothesis~\eqref{eq:Hcanm} to give
\begin{equation}\label{eq:Hcan1}
\mathcal{H}_{\mathrm{can}\,1}^{\mu\nu}\bigl[\dM_{\leqslant n-1\,L}, \dS_{\leqslant n-1\,L}\bigr] + G^{n-1} \mathcal{H}_{\mathrm{can}\,1}^{\mu\nu}\bigl[\dM_{n\,L}, \dS_{n\,L}\bigr] = \mathcal{H}_{\mathrm{can}\,1}^{\mu\nu}\bigl[\dM_{\leqslant n\,L}, \dS_{\leqslant n\,L}\bigr]\,.
\end{equation}
Finally, in any of the non-linear approximations in our induction hypothesis, \textit{i.e.} Eq.~\eqref{eq:Hcanm} with $m\geqslant 2$, we are entitled to consistently replace the previous set of moments $\{\dM_{\leqslant n-1\,L},$ $\dS_{\leqslant n-1\,L}\}$ by the more accurate, newly determined set $\{\dM_{\leqslant n\,L}, \dS_{\leqslant n\,L}\}$, modulo higher-order PM terms at least of order $\propto G^{n+1}$, which we can discard to order $n$. Hence we have proved that the $n$PM contribution to the canonical moments $\{\dM_L, \dS_L\}$ is determined, as is the $n$PM piece of the coordinate shift $\varphi^\mu$, and our recurrence hypothesis is verified at the next order $n$. The practical implementation at quadratic and cubic orders is described in Sec.~\ref{sec:implementation}. As a verification, we have also followed an alternative approach, presented in App.~\ref{app:method_2}.

\subsection{Extraction of physical multipole moments}

A straightforward procedure permits reading off the expressions of the $n$-th order corrections to the canonical moments $\{\dM_{n\,L}, \dS_{n\,L}\}$ from Eq.~\eqref{eq:Hn}, where the left-hand side $H^{\mu\nu}_{n}=U^{\mu\nu}_{n}+V^{\mu\nu}_{n}$ follows from the algorithm~\eqref{eq:hgenericn}. In fact, as we have seen that $U_n^{\mu\nu} = X_n^{\mu\nu} + Y_n^{\mu\nu}$ is a retarded vacuum solution of the wave equation, we can directly, and uniquely, give the desired moments as functions of the ten sets of retarded STF moments composing $U_n^{\mu\nu}$. We thus pose
\begin{subequations}\label{eq:XnYnexpandedSTF}
	\begin{align}
	U_n^{00} &=
	\sum_{\ell \geqslant 0}\,\partial_L\left[ \frac{1}{r} A_L\left(t- \frac{r}{c}\right) \right]\,,\\
	U_n^{0i} &=
	\sum_{\ell\geqslant 0}\,\partial_{iL} \left[ \frac{1}{r} B_L \left(t- \frac{r}{c}\right)\right]
	+ \sum_{\ell\geqslant 1}\,\partial_{L-1} \left[ \frac{1}{r} C_{iL-1} \left(t- \frac{r}{c}\right)\right]
	\nonumber\\ & + \sum_{\ell\geqslant 1}\,\partial_L \left[ \frac{1}{r} D_{i\vert L} \left(t- \frac{r}{c}\right)\right]
	\,,\\
	U_n^{ij} &=
	\sum_{\ell\geqslant 0}\,\partial_{ijL} \left[ \frac{1}{r} E_L \left(t- \frac{r}{c}\right)\right]
	+ \sum_{\ell\geqslant 0}\,\delta_{ij}\,\partial_{L} \left[ \frac{1}{r} F_L \left(t- \frac{r}{c}\right)\right]
	+ \sum_{\ell\geqslant 1}\,\partial_{L-1(i} \left[ \frac{1}{r} G_{j)L-1} \left(t- \frac{r}{c}\right)\right]
	\nonumber\\ &+ \sum_{\ell\geqslant 1}\,\partial_{L(i}\left[ \frac{1}{r} H_{j)\vert L} \left(t- \frac{r}{c}\right)\right]
	+ \sum_{\ell\geqslant 2}\,\partial_{L-2} \left[ \frac{1}{r} K_{ijL-2} \left(t- \frac{r}{c}\right)\right]
	\nonumber\\ & + \sum_{\ell\geqslant 2}\,\partial_{L-1} \left[ \frac{1}{r} L_{(i\vert j)L-1} \left(t- \frac{r}{c}\right)\right]\,,
	\end{align}
\end{subequations}
and follow the steps~\eqref{eq:hgenericn}, successively computing $W_n^{\mu}$, $V_n^{\mu\nu}$ and $H_n^{\mu\nu}$, which is finally put into the form~\eqref{eq:Hn} on which we read off the physical moments
\begin{subequations}\label{eq:dMnL}
\begin{align}
\dM_{n\,L} &= - \frac{c^2(-)^\ell \ell!}{4}\biggl[A_{L}+4\,\frac{B^{(1)}_{L}}{c}+3\,\frac{E^{(2)}_{L}}{c^2}+3F_{L}+G_{L}\biggr]\,,&(\ell\geqslant 2)\label{eq:dMnLa}\\
\dS_{n\,L} &= \frac{c^3(-)^\ell (\ell+1)!}{4\ell}\biggl[D_{L}+\frac{H^{(1)}_{L}}{2c}\biggr]\,.&(\ell\geqslant 2)
\end{align}
\end{subequations}
For completeness, we also give the corrections to the gauge moments composing the gauge vector as $\psi_n^{\mu}$:
\begin{subequations}\label{eq:dWnL}
\begin{align}
\dW_{n\,L} &=
	\frac{c^3(-)^\ell\,\ell !}{4}\biggl[B_L+\frac{E^{(1)}_L}{2c}\biggr]\,,&(\ell\geqslant 0)\\
\dX_{n\,L} &= 
	- \frac{c^4(-)^\ell\,\ell !}{8}\,E_L\,,&(\ell\geqslant 0)\\
\dY_{n\,L} &= 
	\frac{3c^4(-)^\ell\,\ell !}{4}\biggl[\frac{B^{(1)}_L}{c}+\frac{E^{(2)}_L}{c^2}+F_L+\frac{G_L}{3}\biggr]\,,&(\ell\geqslant 1)\\
\dZ_{n\,L} &= 
	- \frac{c^4(-)^\ell\,(\ell+1)!}{8\ell}\,H_L\,.&(\ell\geqslant 1)
	\end{align}
\end{subequations}

\section{Practical implementation}
\label{sec:implementation}

Let us apply the previously described procedure to determine the quadratic and cubic corrections $\mathcal{M}_{2\,ij}$ and $\mathcal{M}_{3\,ij}$ to the mass-type quadrupole moment $\dM_{ij}$ at the 4PN order [see Eqs.~\eqref{eq:ML_SL}]. Previous investigations~\cite{B96, BFIS08} focused on quadratic interactions and determined the mass quadrupole at leading order 2.5PN and sub-leading order 3.5PN; such corrections are recalled in Eq.~\eqref{eq:resMij}. However, in order to obtain the 4PN correction, we need to derive the cubic interactions. A preliminary dimensional analysis shows that at 4PN order and in the center-of-mass frame (where the mass dipole $\dI_i$ is vanishing), the only multipole interactions between the source and gauge moments are cubic and necessarily of the three types:
\begin{align}\label{eq:cubicinter}
\dM \times \dM \times \dW_{ij}\,,\qquad \dM \times \dM \times \dY_{ij}\,,\qquad \dM \times \dI_{ij} \times \dW\,, 
\end{align}
where $M$ is the constant ADM mass, $\dI_{ij}$ is the source mass quadrupole moment, and where the monopole $\dW$ and the two quadrupoles $\dW_{ij}$ and $\dY_{ij}$ are gauge moments. Note that out of the center-of-mass frame, a large number of additional interactions would appear, such as $\dI_i\times\dI_i\times\dW$, $\dM\times\dI_i\times\dW_i$, $\dots$, but those are not needed in concrete applications. 

\subsection{Controlling the cubic source terms}
\label{sec:controlling}

The first step towards the practical determination of the cubic couplings is to successively construct the quadratic and cubic quantities $\Omega_n^{\mu\nu}$ and $\Delta_n^\mu$ that enter Eqs.~\eqref{eq:commXY} for $n=2,3$. To quadratic order we have
\begin{subequations}\label{eq:OmegaDelta}
	\begin{align}
	\label{eq:Omega}	
	\Omega_{2}^{\mu\nu}
	\equiv
	& - \partial_\rho\left[\varphi_1^\rho\left(h_{\text{can}\,1}^{\mu\nu}+\partial\varphi_1^{\mu\nu}\right)\right]
	+ 2\,\partial_\rho\varphi_1^{(\mu}\,h_{\text{can}\,1}^{\nu)\rho}\\
	& 
	+ \partial^\rho\varphi_1^{(\mu}\,\partial_\rho\varphi_1^{\nu)}
	+\frac{1}{2}\eta^{\mu\nu}\left[\partial_\rho\varphi_1^\sigma\partial_\sigma\varphi_1^\rho-\partial_\rho\varphi_1^\rho\partial_\sigma\varphi_1^\sigma\right]\,,\nonumber
	\\
	\label{eq:Delta}	
	\Delta_{2}^\mu 
	\equiv 
	& - h_{\text{can}\,1}^{\rho\sigma}\,\partial_{\rho\sigma}\varphi_1^\mu\,.
	%	\label{eq:Delta}
	\end{align}
\end{subequations}
This is valid ``on-shell'', as we have used the facts that $\Box h^{\mu\nu}_{\text{can}\,1}  = \Box \varphi_1^\mu =0$, which hold at linear order. One can directly verify that [see Eq.~\eqref{eq:divordren}]
\begin{equation}\label{eq:relationOmegaDelta}
\partial_\nu\Omega^{\mu\nu}_{2} + \Delta^\mu_{2} = 0\,.
\end{equation}

At the linear level, as is clear from Eqs.~\eqref{eq:h1can}--\eqref{eq:phi1}, $\dM$ and $\dI_{ij}$ enter $h_{\text{can}\,1}^{\mu\nu}$ whereas $\dW$, $\dW_{ij}$ and $\dY_{ij}$ enter $\varphi_1^{\mu}$. Thus, before tackling the cubic interactions, we naturally require the knowledge of the complete quadratic interactions $\dM\times\dM$, $\dM\times\dI_{ij}$, $\dM\times\dW$, \emph{etc}. The canonical quadratic metrics $h_{\text{can}\,1}^{\mu\nu}$ corresponding to the interactions $\dM\times\dM$ and $\dM\times\dI_{ij}$ can be found in~\cite{BD92, B98tail}. The quadratic couplings between source and gauge moments have been computed in~\cite{B96, BFIS08}. Notably, this yields [see~\eqref{eq:dMnLa}]
\begin{equation}
\dM_{2\,ij} = \frac{4}{c^5}\left[\dW^{(2)}\dI_{ij} - \dW^{(1)}\dI_{ij}^{(1)}\right] + \calO\left(\frac{1}{c^7}\right)\,,
\end{equation}
and we have already given the 3.5PN contribution in~\eqref{eq:resMij}.

What remains to be computed, for insertion into the cubic quantities $\Omega_3^{\mu\nu}$ and $\Delta_3^\mu$, are the quadratic couplings contributing to the quadratic coordinate shift $\varphi_2^\mu$. As we have seen in~\eqref{eq:varphin}, the shift is composed of two parts: the first one, $\phi_2^\mu$, has been defined in Eq.~\eqref{eq:phin}, and is computed by the usual techniques (see for instance the Appendix A of~\cite{B98quad}). The second part, $\psi_2^\mu$, is extracted from~\eqref{eq:Hn} as quadratic corrections to the four types of gauge moments; the general result has been provided in Eqs.~\eqref{eq:dWnL}. Working out $\varphi_2^\mu=\phi_2^\mu+\psi_2^\mu$ for the needed quadratic interactions $\dM\times\dW$, $\dM\times\dW_{ij}$, $\dM\times\dY_{ij}$ and $\dI_{ij}\times\dW$, we find
\begin{subequations}\label{eq:varphiexpl}
\begin{align}
\varphi^0_{\dM\times\dW} &=  -\frac{16\dM}{c^7}\int_1^{+\infty}\!\!\dd y\,Q_0(y)\dW^{(2)}\left(t-\frac{yr}{c}\right)\,, \\
\varphi^i_{\dM\times\dW} &= 0\,, \\[0.4cm]
%%%%%%%%%%%%%%%%%%%%%%%%%%%%%%%%%%%%%%%%%%%%%%%%%%%%%%%%%%%%%%%%
\varphi^0_{\dM\times\dW_{ij}} &= -\frac{6\dM}{c^7\,r^2}\hat{n}^{ij}\biggl[\dW^{(2)}_{ij}\left(t-\frac{r}{c}\right)+\frac{r}{c}\,\dW^{(3)}_{ij}\left(t-\frac{r}{c}\right) \\
&  \qquad\qquad\qquad +\frac{4r^2}{3c^2}\int_1^{+\infty}\!\!\dd y\,Q_2(y)\dW^{(4)}_{ij}\left(t-\frac{yr}{c}\right)\biggr]\,,\nonumber\\
\varphi^i_{\dM\times\dW_{ij}} &= 0\,,\\[0.4cm]
%%%%%%%%%%%%%%%%%%%%%%%%%%%%%%%%%%%%%%%%%%%%%%%%%%%%%%%%%%%%%%%%
\varphi^0_{\dM\times\dY_{ij}} &=0\,,\\ 
\varphi^i_{\dM\times\dY_{ij}} &=-\frac{4\dM}{c^8\,r}\hat{n}^j\left[\dY^{(2)}_{ij}\left(t-\frac{r}{c}\right)+\frac{2r}{c}\int_1^{+\infty}\!\!\dd y\,Q_1(y)\dY^{(3)}_{ij}\left(t-\frac{yr}{c}\right)\right]\,,\\[0.4cm]
%%%%%%%%%%%%%%%%%%%%%%%%%%%%%%%%%%%%%%%%%%%%%%%%%%%%%%%%%%%%%%%%
\varphi^0_{\dI_{ij}\times \dW} &= - \frac{2}{c^8\,r}\hat{n}^{ij}\bigg[\dI_{ij}\dW^{(3)}+\dI_{ij}^{(1)}\dW^{(2)}+\dI_{ij}^{(2)}\dW^{(1)}+\dI_{ij}^{(3)}\dW\bigg]\\
&\quad \nonumber
- \frac{6}{c^7\,r^2}\hat{n}^{ij}\bigg[\dI_{ij}\dW^{(2)}-\frac{2}{3}\dI_{ij}^{(1)}\dW^{(1)}+\dI_{ij}^{(2)}\dW\bigg]\,,\\
\varphi^i_{\dI_{ij}\times \dW} &= 0\,. 
\end{align}
\end{subequations}
Note the presence of non-local tail integrals, involving $Q_m(y)$, the Legendre functions of the second kind, here defined with a branch cut from $-\infty$ to 1 and related to the Legendre polynomials $P_m(y)$ by
\begin{equation}\label{eq:defQm}
	Q_{m} (y) = \frac{1}{2} P_m (y) \, \mathrm{ln}\left(\frac{y+1}{y-1} \right)- \sum^{m}_{ j=1} \frac{1}{j} P_{m-j}(y) P_{j-1}(y)\,.
\end{equation}
We inject the expressions~\eqref{eq:varphiexpl} together with the expressions for the canonical metric, notably $h^{\mu\nu}_{\text{can}\,2}$ corresponding to the interaction $\dM\times\dI_{ij}$ which is also non-local and given by~(B3) in~\cite{BD92}, into the cubic $\Omega_3^{\mu\nu}$ and $\Delta_3^\mu$ that define~\eqref{eq:commXY}. For convenience, we split $\Omega_{3}^{\mu\nu}$ into quadratic-type and purely cubic interactions: $\Omega_{3}^{\mu\nu} \equiv \Omega_{12}^{\mu\nu} + \Omega_{21}^{\mu\nu} + \Omega_{111}^{\mu\nu}$, where
\begin{subequations}\label{eq:Omega3}
	\begin{align}
	\Omega_{12}^{\mu\nu} =& - \partial_\rho\left[\varphi_1^\rho\left(h_{\text{can}\,2}^{\mu\nu}+\partial\varphi_2^{\mu\nu}\right)\right]
	+ 2\,\partial_\rho\varphi_1^{(\mu}\,h_{\text{can}\,2}^{\nu)\rho}\nonumber\\
	& 
	+ \partial^\rho\varphi_1^{(\mu}\,\partial_\rho\varphi_2^{\nu)}
	+\frac{1}{2}\eta^{\mu\nu}\left[\partial_\rho\varphi_1^\sigma\partial_\sigma\varphi_2^\rho-\partial_\rho\varphi_1^\rho\partial_\sigma\varphi_2^\sigma\right]\,,\\
	%%%%%%%%%%%%%%%%%%%%%%%%%%%%%%%%%%%%%%%%%%%%%%%%%%%%%%%%%%%%%%%
	\Omega_{21}^{\mu\nu} =& - \partial_\rho\left[\varphi_2^\rho\left(h_{\text{can}\,1}^{\mu\nu}+\partial\varphi_1^{\mu\nu}\right)\right]
	+ 2\,\partial_\rho\varphi_2^{(\mu}\,h_{\text{can}\,1}^{\nu)\rho}\nonumber\\
	& 
	+ \partial^\rho\varphi_2^{(\mu}\,\partial_\rho\varphi_1^{\nu)}
	+\frac{1}{2}\eta^{\mu\nu}\left[\partial_\rho\varphi_2^\sigma\partial_\sigma\varphi_1^\rho-\partial_\rho\varphi_2^\rho\partial_\sigma\varphi_1^\sigma\right]\,,\\
	%%%%%%%%%%%%%%%%%%%%%%%%%%%%%%%%%%%%%%%%%%%%%%%%%%%%%%%%%%%%%%%
	\Omega_{111}^{\mu\nu} =& \frac{1}{2}\,\partial_{\rho\sigma}\left(\varphi_1^\rho\varphi_1^\sigma\,h_{\text{can}\,1}^{\mu\nu}\right)
	+ h_{\text{can}\,1}^{\rho\sigma}\,\partial_\rho\varphi_1^\mu\,\partial_\sigma\varphi_1^\nu
	-2 \partial_\rho\left(\varphi_1^\rho\, h_{\text{can}\,1}^{\sigma(\mu}\,\partial_\sigma\varphi_1^{\nu)}\right)\nonumber\\
	& 
	+ \frac{1}{2}\,\partial_{\rho\sigma}\left(\varphi_1^\rho\varphi_1^\sigma\,\partial\varphi_1^{\mu\nu}\right)
	-\partial_\rho\left(\varphi_1^\rho\,\partial_\sigma\varphi_1^\mu\,\partial^\sigma\varphi_1^\nu\right)
	- \frac{1}{2}\eta^{\mu\nu}\,\varphi_1^\lambda\partial_\lambda\left(\partial_\rho\varphi_1^\sigma\partial_\sigma\varphi_1^\rho-\partial_\rho\varphi_1^\rho\partial_\sigma\varphi_1^\sigma\right)
	\nonumber\\
	& 
	+ \frac{1}{3}\eta^{\mu\nu}\left(\partial_\rho\varphi_1^\rho\partial_\sigma\varphi_1^\sigma\partial_\lambda\varphi_1^\lambda-\partial_\rho\varphi_1^\sigma\partial_\sigma\varphi_1^\lambda\partial_\lambda\varphi_1^\rho\right)\,.
	\end{align}
\end{subequations}
Recall that our computations are done on-shell, using the wave equations satisfied at linear order, $\Box h_{\text{can}\, 1}^{\mu\nu} = \Box \varphi_1^{\mu} = 0$, and quadratic order, $\Box h_{\text{can}\, 2}^{\mu\nu} = \Lambda_{\text{can}\, 2}^{\mu\nu}$. An important point is that $\Omega_{111}^{\mu\nu}$ only contains $h_{\text{can}\,1}\times\varphi_1\times\varphi_1$ and $\varphi_1\times\varphi_1\times\varphi_1$ terms, but there is no $h_{\text{can}\,1}\times h_{\text{can}\,1}\times\varphi_1$ sector. This fact allows us to discard it entirely in the practical implementation, since at the 4PN order we have only the three cubic interactions~\eqref{eq:cubicinter} which contain at most one gauge moment. Similarly, we have for the coordinate shift $\Delta_{3}^{\mu} \equiv \Delta_{12}^{\mu} + \Delta_{21}^{\mu} + \Delta_{111}^\mu$ with
\begin{subequations}\label{eq:Delta3}
	\begin{align}
	\Delta_{12}^{\mu} &= - h_{\text{can}\,2}^{\rho\sigma}\,\partial_{\rho\sigma}\varphi_1^\mu \,,\\[0.2cm]
	%%%%%%%%%%%%%%%%%%%%%%%%%%%%%%%%%%%%%%%%%%%%%%%%%%%%%%%%%%%%%%%
	\Delta_{21}^{\mu} &= - h_{\text{can}\,1}^{\rho\sigma}\,\partial_{\rho\sigma}\varphi_2^\mu + \partial_\rho\left(\varphi_1^\rho \Delta_2^\mu\right)\,,\\[0.2cm]
	%%%%%%%%%%%%%%%%%%%%%%%%%%%%%%%%%%%%%%%%%%%%%%%%%%%%%%%%%%%%%%%
	\Delta_{111}^{\mu} &= \partial_\rho\left(\varphi_1^\rho\,h_{\text{can}\,1}^{\rho\sigma}\partial_{\rho\sigma}\varphi_1^\mu\right)\,.
	\end{align}
\end{subequations}
We have inserted $\Box \varphi_2^\mu=\Delta_2^\mu$, which comes from~\eqref{eq:phin}, and used the fact that $\Box \psi_2^\mu=0$. The latter quantities satisfy $\partial_\nu\Omega_{3}^{\mu\nu}+\Delta_{3}^\mu=0$, consistently with the divergencelessness of the cubic source, and in fact, even separately,
\begin{equation}
\partial_\nu\Omega_{12}^{\mu\nu}+\Delta_{12}^\mu=0\,,\qquad\partial_\nu\Omega_{21}^{\mu\nu}+\Delta_{21}^\mu=0\,,\qquad\partial_\nu\Omega_{111}^{\mu\nu}+\Delta_{111}^\mu=0\,.
\end{equation}

\subsection{Retarded integral of the source terms}

\label{sec:retarded_integrals}

At this stage, we control the cubic source terms that are required to evaluate the ``commutators'' $X_3^{\mu\nu}$ and $Y_3^{\mu\nu}$ defined by the formulas~\eqref{eq:commXY}. Using canonical relations between the Legendre functions, we find that the terms to be computed fall into two and only two classes:
\begin{subequations}\label{eq:IinstItail}
	\begin{align}
	& 
	\mathcal{I}_\text{inst}
	\equiv \FPprop \biggl[\left(\frac{r}{r_0}\right)^B\! B^b\,\frac{\hat{n}_L}{r^p}\,F\left(t-\frac{r}{c}\right)\biggr]\,,\label{eq:I1}\\[0.2cm]
	& 
	\mathcal{I}_\text{tail}
	\equiv \FPprop \biggl[\left(\frac{r}{r_0}\right)^B\! B^b\,\frac{\hat{n}_L}{r^p}\,G\left(t-\frac{r}{c}\right)\int_1^{+\infty}\!\!\dd y\,Q_m(y)\,F\left(t-\frac{yr}{c}\right)\biggr]\,.\label{eq:I2}
	\end{align}
\end{subequations}
Here, the functions $F$ and $G$ represent some products of (source or gauge) multipole moments, with the function $G$ which can be constant when dealing with the ADM mass. The second type of term $\mathcal{I}_\text{tail}$ integrates over a tail term which comes from the tail present in the quadratic metric for the interaction $\dM\times\dI_{ij}$, see~(B3) in~\cite{BD92}, and those which appeared in the coordinate shift [see Eqs.~\eqref{eq:varphiexpl}]. The crucial point here is the presence of the explicit factors $B$ and $B^2$ in the definitions~\eqref{eq:commXY} hence we posed $b=1, 2$ in Eqs.~\eqref{eq:IinstItail}; $p$ in an integer and the index of the Legendre function is $m \in [0,4]$; and $\hat{n}_L$ is the usual STF harmonics.

Thanks to the factors $B$ or $B^2$, we know that the integrals~\eqref{eq:IinstItail} will depend only on the behaviour of the source terms when $r\to 0$.\footnote{Recall that in the MPM approach the $r\to +\infty$ limit is harmless as we assume that the multipole moments, \textit{i.e.} the functions $F$ and $G$, become stationary in the remote past.} Indeed the regularization process $\FP$ has been introduced to cope with the singular behaviour of the source when $r\to 0$, and hence only that limit can generate poles $1/B$ or $1/B^2$ which can compensate the factors $B$, $B^2$ and lead to a finite part when $B\to 0$. Therefore we are entitled to restrict ourselves to a ball of finite and even infinitesimal size (say $r<\epsilon$) and replace the source terms in~\eqref{eq:IinstItail} by their formal Taylor expansion when $r\to 0$ (\textit{i.e.}, the near zone or PN expansion $r/c\to 0$). Once the near zone expansion of the source is known the integration can be performed with standard techniques. The result for the simpler case $\mathcal{I}_\text{inst}$ (with purely instantaneous source term) was already given in Eq.~(A.18) of~\cite{B98quad}. It is non-zero only when $p\geqslant\ell+3$ and $b=1$ (only a simple pole $1/B$ can appear in this case), and reads
\begin{equation}\label{eq:resI1}
\mathcal{I}_\text{inst} = \left\{\begin{array}{ll} \displaystyle  \frac{(-)^{p}2^{p-3}(p-3)!}{(p+\ell-2)!(p-\ell-3)!c^{p-\ell-3}}\,\hat{\partial}_L\left[\frac{F^{(p-\ell-3)}(t-r/c)}{r}\right] &\quad\text{($b=1$ and $p\geqslant\ell+3$)}\,,\\[0.6cm]\displaystyle 
	0 &\quad\text{($b=2$ or $p<\ell+3$)}\,.\end{array}\right.
\end{equation}

Next, we deal with the more difficult integral $\mathcal{I}_\text{tail}$. By the previous argument we can replace the function $G(t-r/c)$ by its formal Taylor expansion when $r\to 0$. The main problem is therefore the control of the expansion series when $r\to 0$ of the tail integral $\mathcal{F}_m  \equiv \int_1^{+\infty}\!\!\dd y \,Q_m(y)\,F(t-y r/c)$. We devote the App.~\ref{app:NZ_tail} to this not-so-easy question and state here the general result for this expansion:
\begin{align}\label{eq:resFmfinal}
\mathcal{F}_m 
& \stackrel{r/c\to 0}{~=~} \,\sum_{i=0}^{+\infty} \beta_{i}^m\,\frac{(-)^i}{i!}\left(\frac{r}{c}\right)^i F^{(i)}(t)\\
&\qquad + \sum_{j=0}^{+\infty} \frac{(-)^m\,c_{j}^m}{(m+2j)!}\left(\frac{r}{c}\right)^{m+2j}\int_0^{+\infty}\!\!\dd\tau\biggl[\ln\left(\frac{c\tau}{2r}\right) - H_{m+j} + 2 H_{2m+2j+1} \biggr]\, F^{(m+2j+1)}\left(t-\tau\right)\,.\nonumber
\end{align}
Here, the symbol $\stackrel{r/c\to 0}{~=~}$ indicates the formal asymptotic expansion, the coefficients $c_j^m$ and $\beta_i^m$ are defined in~\eqref{eq:cmj} and~\eqref{eq:betami}, and $H_n$ denotes the usual harmonic number. Thus, we have to multiply Eq.~\eqref{eq:resFmfinal} by the Taylor expansion of $G(t-r/c)$ and integrate term by term. Note that all those formal expansions are convergent, due to the stationary of the source in the remote past. As we see from the structure of the expansion~\eqref{eq:resFmfinal}, the final integration will boil down to the control of just one type of term,
\begin{equation}\label{eq:J}
	\mathcal{J}
	\equiv \FPprop \biggl[\left(\frac{r}{r_0}\right)^B\! B^b\,\frac{\hat{n}_L}{r^p}\,(\ln r)^a\,H(t)\biggr]\,.
\end{equation}
However, the function $H(t)$ here can be either an instantaneous function or a non-local tail integral of the type given in~\eqref{eq:resFmfinal}. Furthermore, still because of the tail term we must add the case where there is logarithm $\ln r$ in the source, see again~\eqref{eq:resFmfinal}. Thus, we consider~\eqref{eq:J} with $a=0$ or $1$. 

We give a few details on the calculation of $\mathcal{J}$. Writing~\eqref{eq:J} in ordinary three-dimensional form we have, since as we said the integration is limited to an infinitesimal ball $r<\epsilon$,
\begin{equation}\label{eq:J3d}
\mathcal{J} =  - \frac{1}{4\pi}\FP \int_{r<\epsilon}\frac{\dd^3\mathbf{x}'}{\vert\mathbf{x}-\mathbf{x}'\vert}\left(\frac{r'}{r_0}\right)^B B^b\,\frac{\hat{n}'_L}{r'^p}\,(\ln r')^a\,H\left(t-\frac{\vert\mathbf{x}-\mathbf{x}'\vert}{c}\right)\,.
\end{equation}
Using the formal STF expansion of the multipolar factor when $r'\equiv\vert\mathbf{x}'\vert\to 0$,
\begin{equation}
\frac{H(t-\vert \mathbf{x}-\mathbf{x}'\vert/c)}{\vert\mathbf{x}-\mathbf{x}'\vert} 
= \sum_{q=0}^{+\infty}\frac{(-)^q}{q!}\sum_{j=0}^{+\infty}\,\frac{r'^{2j}}{2^j j!}\,\frac{(2q+1)!!}{(2q+2j+1)!!}\,\hat{x}'_Q\hat{\partial}_Q\left(\frac{H^{(2j)}(t-r/c)}{r c^{2j}}\right)\,,
\end{equation}
together with the angular integration performed using Eq.~(A29a) in~\cite{BD86}, we end up with radial integrals of the type
\begin{equation}
\FP \int_0^\epsilon \dd r' \,\left(\frac{r'}{r_0}\right)^B B^b\,r'^{\ell+2j+2-p}\,(\ln r')^a = \FP \frac{B^b}{r_0^B}\left(\frac{\dd}{\dd B}\right)^a\left[\frac{\epsilon^{B+\ell+2j+3-p}}{B+\ell+2j+3-p}\right] \,.
\end{equation}
Here the $r'=0$ boundary vanishes by analytic continuation on $B$. With the factor $B^b$ in front the latter integral is zero unless there is a pole, and the pole can come only when $p$ is of the form $p=\ell+3+2j$ (where $j\in\mathbb{N}$). Hence the results are immediate. 

When $a=0$ (source without $\ln r$) the integral is zero when $b=2$ as there is only a simple pole $1/B$ in this case:
\begin{subequations}\label{eq:resJ}
\begin{equation}\label{eq:resJa=0}
\mathcal{J}_{a=0} = \left\{\begin{array}{ll} \displaystyle  \frac{(-)^{p}}{(p+\ell-2)!!(p-\ell-3)!!c^{p-\ell-3}}\,\hat{\partial}_L\left[\frac{H^{(p-\ell-3)}(t-r/c)}{r}\right] &~\text{($b=1$ and $p-\ell-3\in 2\mathbb{N}$)}\,,\\[0.6cm]\displaystyle 
0 &~\text{($b=2$ or $p-\ell-3\notin2\mathbb{N}$)}\,.\end{array}\right.
\end{equation}
One can check that this is perfectly consistent with the result for $\mathcal{I}_\text{inst}$ in~\eqref{eq:resI1}, in the sense that one can Taylor expand the source term of $\mathcal{I}_\text{inst}$ when $r\to 0$ and recover the same result by applying Eq.~\eqref{eq:resJa=0} on each term of the Taylor series. By contrast, when there is a $\ln r$ the integral is also non-zero when $b=2$ because of the presence of a double pole $1/B^2$:
\begin{equation}
\mathcal{J}_{a=1} = \left\{\begin{array}{ll} \displaystyle 
\frac{(-)^p\,\ln r_0}{(p+\ell-2)!!(p-\ell-3)!!c^{p-\ell-3}}\,\hat{\partial}_L\left[\frac{H^{(p-\ell-3)}(t-r/c)}{r}\right] &~\text{($b=1$ and $p-\ell-3\in 2\mathbb{N}$)}\,,
\\[0.4cm]\displaystyle 
\frac{(-)^{p+1}}{(p+\ell-2)!!(p-\ell-3)!!c^{p-\ell-3}}\,\hat{\partial}_L\left[\frac{H^{(p-\ell-3)}(t-r/c)}{r}\right] &~\text{($b=2$ and $p-\ell-3\in 2\mathbb{N}$)}\,,
\\[0.6cm] \displaystyle  0 &~\text{($p-\ell-3\notin 2\mathbb{N}$)}\,.
\end{array}\right.
\end{equation}
\end{subequations}

Finally with those results in hand, we can implement all the terms up to cubic non-linear order, using the \emph{xAct} library of the \emph{Mathematica} software~\cite{xtensor}. Such computation ends up with the final result at the 4PN level already recapitulated in Eq.~\eqref{eq:resMij}. In the App.~\ref{app:method_2}, an alternative procedure is described which permitted checking it independently. 

\acknowledgments

We acknowledge discussions with Laura Bernard, Quentin Henry and David Trestini. F.L. received funding from the European Research Council (ERC) under the European Union's Horizon 2020 research and innovation program (grant agreement No 817791).

\appendix

\section{Near-zone expansion of the tail integral}\label{app:NZ_tail}

This appendix is devoted to the determination of the near zone (or PN) expansion when $r/c\to 0$ of the tail integral that enters the source of the integral $\mathcal{I}_\text{tail}$~\eqref{eq:I2}. This will permit justifying the claim of Eq.~\eqref{eq:resFmfinal} concerning the structure of the near zone expansion, and provide the explicit coefficients we need for the practical computation. Thus we look for the expansion of 
\begin{equation}\label{eq:defFm}
\mathcal{F}_m(r,t) \equiv \int_1^{+\infty}\!\!\dd y \,Q_m(y)\,F\left(t-\frac{yr}{c}\right)\,.
\end{equation}
If an explicit expression of the Legendre function of the second kind $Q_m(y)$ is given in Eq.~\eqref{eq:defQm}, we preferably use here the general expression of the Legendre function for generic $\mu\in\mathbb{C}\setminus \{-1,-2,\cdots\}$ in terms of the hypergeometric function $F\equiv{}_2F_1$:
\begin{equation}\label{eq:Qmu}
Q_\mu (y) = \frac{\sqrt{\pi}}{2^{\mu+1}}\frac{\Gamma\left(\mu+1\right)}{\Gamma\left(\mu+\frac{3}{2}\right)} \,\frac{1}{y^{\mu+1}} \,F\left(\frac{\mu}{2}+1, \frac{\mu+1}{2}; \mu+\frac{3}{2}; \frac{1}{y^2}\right)\,,
\end{equation}
with $\vert y\vert>1$, $\vert\arg y\vert<\pi$. This representation of the Legendre function reads explicitly, for any real argument such that $y>1$,
\begin{subequations}\label{eq:Qmuexpl}
	\begin{align}
	Q_\mu (y) &= \sum_{j=0}^{+\infty} c_j^\mu\, y^{-1-\mu-2j}\,,\\
	\text{with}\quad c_j^\mu &\equiv \frac{2^\mu}{j!}\frac{\Gamma\left(\mu+2j+1\right)\Gamma\left(\mu+j+1\right)}{\Gamma\left(2\mu+2j+2\right)}\,.\label{eq:cjmu}
	\end{align}
\end{subequations}
It can naturally be regarded as an asymptotic expansion when $y\to+\infty$ on the real axis. However, we stress that it is valid as soon as $y\in ]1,+\infty[$, as the series representation of the hypergeometric function is well defined in that case.

Defining $\tau \equiv yr/c$ one can then rewrite~\eqref{eq:defFm} as the following series
\begin{subequations}\label{eq:Fr0}
	\begin{align}
	\mathcal{F}_m &= \sum_{j = 0}^{+\infty}\,c_{j}^m\, \mathcal{F}_{m}^j\,,\\
	\text{where}\quad \mathcal{F}_{m}^j &\equiv 
	\left(\frac{r}{c}\right)^{m+2j}\int_{r/c}^{+\infty}\!\!\dd\tau\,\tau^{-m-2j-1}\,F(t-\tau)\,,
	\end{align}
\end{subequations}
and the coefficients are just given by~\eqref{eq:cjmu} in the particular case where $m\in\mathbb{N}$, \textit{i.e.},
\begin{equation}\label{eq:cmj}
c_{j}^m = \frac{2^m}{j!}\frac{(m+2j)!(m+j)!}{(2m+2j+1)!}\,.
\end{equation}
Upon the expression of $\mathcal{F}_{m}^j$ we perform a series of integrations by parts in order to increase the power of $\tau$ until we reach a logarithm of $\tau$. The all-integrated terms contain the functions $F^{(n)}(t-r/c)$ that we replace by their Taylor expansions when $r\to 0$. The last integral containing $\ln\tau$ is split according to $\int_{r/c}^{+\infty}=-\int_0^{r/c}+\int_0^{+\infty}$. In the first integral, from $0$ to $r/c$, we are allowed to expand the integrand when $\tau\to 0$, since by definition $r/c\to 0$ for the PN expansion. Then the second integral from $0$ to $+\infty$ just gives a tail integral in the ordinary sense. Therefore, we obtain the following asymptotic expansion when $r/c\to 0$:
\begin{align}\label{eq:Fmj}
  \mathcal{F}_{m}^j &\stackrel{r/c\to 0}{~=~} \sum_{\substack{i=0\\i\neq m+2j}}^{+\infty} \frac{(-)^i}{i!\,(m+2j-i)}\left(\frac{r}{c}\right)^i F^{(i)}\left(t\right)
\nonumber\\
& + \frac{(-)^m}{(m+2j)!}\left(\frac{r}{c}\right)^{m+2j}\int_0^{+\infty}\!\!\dd\tau\biggl[\ln\left(\frac{c\tau}{r}\right) + H_{m+2j} \biggr]\, F^{(m+2j+1)}\left(t-\tau\right)\,.
\end{align}
Very importantly, the value $i=m+2j$ is to be excluded from the first summation. The last term is the tail integral, where $H_n$ denotes the usual harmonic number. Resumming on $j$ yields 
\begin{align}\label{eq:resFm}
& \mathcal{F}_m \stackrel{r/c\to 0}{~=~} \sum_{i=0}^{+\infty} \alpha_{i}^m\,\frac{(-)^i}{i!}\left(\frac{r}{c}\right)^i F^{(i)}(t)
\nonumber\\
&\qquad + \sum_{j=0}^{+\infty} c_{j}^m\,\frac{(-)^m}{(m+2j)!}\left(\frac{r}{c}\right)^{m+2j}\int_0^{+\infty}\!\!\dd\tau\biggl[\ln\left(\frac{c\tau}{r}\right) + H_{m+2j} \biggr]\, F^{(m+2j+1)}\left(t-\tau\right)\,.
\end{align}
We already see the type of structure claimed in Eq.~\eqref{eq:resFmfinal}. The coefficients $\alpha_{i}^m$ in the first sum (corresponding to instantaneous terms) are still at this stage given by an infinite series:
\begin{equation}\label{eq:alphami}
\alpha_{i}^m = \sum_{\substack{j=0\\ i\neq m+2j}}^{+\infty} \frac{c_{j}^m}{m+2j-i}\,.
\end{equation}
Despite that, the result~\eqref{eq:resFm} can be dealt with as it is in practical calculations. However, as it turns out the coefficient~\eqref{eq:alphami} can be resummed in analytic closed form, and this will yield a more interesting and powerful expression of the near zone expansion of $\mathcal{F}_m$.

To obtain such form of $\alpha_{i}^m$ we take advantage of the fact that the coefficients $c_j^m$ can be generalized to any $\mu\in\mathbb{C}\setminus \{-1,-2,\cdots\}$ through the general definition of the Legendre function, see Eqs.~\eqref{eq:Qmuexpl}. Therefore, one can extend the definition of $\alpha_{i}^m$ to any generic value of $\mu$ by posing
\begin{equation}\label{eq:alphaimu}
\alpha_{i}^{\mu} = \sum_{j=0}^{+\infty} \,\frac{c_{j}^{\mu}}{\mu+2j-i}\,,
\end{equation}
where the coefficients $c_{j}^{\mu}$ are now given by~\eqref{eq:cjmu}. We assume that $\mu$ is non-integral, so that the condition $i\neq m+2j$ is no longer necessary and has been dropped. Now, from the very definition of the coefficients $c_{j}^{\mu}$ in~\eqref{eq:Qmuexpl}, the fact that the hypergeometric series is absolutely convergent for $|y|>1$, and using the identity $\int_1^{+\infty} \dd y\, y^i \times y^{-1-\mu-2j}=(\mu+2j-i)^{-1}$ valid for $\Re(\mu)>i$, we prove that the coefficient $\alpha_{i}^\mu$ is actually given by
\begin{equation}\label{eq:alphaQmu}
\alpha_{i}^{\mu} = \int_1^{+\infty}\!\!\dd y \,y^i\,Q_\mu(y)\,.
\end{equation}
Furthermore, we know that~\cite{GR}
\begin{equation}\label{eq:GR}
\int_1^{+\infty} \dd y\, (y-1)^{\nu} \,Q_\mu(y) = 2^{\nu}\frac{[\Gamma(\nu+1)]^2\,\Gamma(\mu-\nu)}{\Gamma(\mu+\nu+2)}\,.
\end{equation}
Hence we arrive at the closed-form expression
\begin{equation}\label{eq:alphaclosedform}
\alpha_{i}^{\mu} = \sum_{k=0}^{i} {\genfrac{(}{)}{0pt}{}{i}{k}}\,2^k(k!)^2\frac{\Gamma(\mu-k)}{\Gamma(\mu+k+2)}\,,
\end{equation}
where ${\genfrac{(}{)}{0pt}{}{i}{k}}$ is the usual binomial coefficient. Its validity may be extended to all non-integral values of $\mu$ by analytic continuation. Still this expression is to be connected to the actual result~\eqref{eq:alphami} we search for, as this result excludes the value $i=m+2j$ from the summation. However, posing $\mu = m+\varepsilon$ one can also substract and then re-add the contributions $i=m+2j$ in~\eqref{eq:alphaimu}; in this way we rewrite $\alpha_{i}^{m}$ as the following limit when $\varepsilon\to 0$:
\begin{equation}\label{eq:alphalimit0}
\alpha_{i}^{m} = \lim_{\varepsilon\to 0}\left\{\begin{array}{ll} \displaystyle   \alpha_{i}^{m+\varepsilon} &\qquad\text{when $i\not= m+2j$}\,,\\[0.6cm]\displaystyle 
\alpha_{m+2j}^{m+\varepsilon} - \frac{1}{\varepsilon}\,c_{j}^{m+\varepsilon} &\qquad\text{when $i = m+2j$}\,.\end{array}\right.
\end{equation}
In the case $i = m+2j$ an explicit pole $1/\varepsilon$ has to be added, and which should cancel the pole present in $\alpha_{m+2j}^{m+\varepsilon}$ so that the limit is finite. Furthermore since the coefficient $c_{j}^{m+\varepsilon}$ is finite when $\varepsilon\to 0$, with limit $c_{j}^{m}$ given by~\eqref{eq:cmj}, we can expand it to order $\varepsilon$ and obtain
\begin{equation}\label{eq:alphalimit}
\alpha_{i}^{m} = \lim_{\varepsilon\to 0}\left\{\begin{array}{ll} \displaystyle   \alpha_{i}^{m+\varepsilon} &\qquad\text{($i\not= m+2j$)}\,,\\[0.4cm]\displaystyle 
\alpha_{m+2j}^{m+\varepsilon} - c_{j}^{m}\Biggl[\frac{1}{\varepsilon} + \left(\frac{\dd\ln c_{j}^{\mu}}{\dd\mu}\right)_{\mu=m}\Biggr] &\qquad\text{($i = m+2j$)}\,.\end{array}\right.
\end{equation}
Finally, the last step is to use the closed-form expression~\eqref{eq:alphaclosedform} for $\alpha_{i}^{m+\varepsilon}$. The extra terms in~\eqref{eq:alphalimit} are easily computed with the help of~\eqref{eq:cjmu} and~\eqref{eq:cmj}. We verify that indeed the limit $\varepsilon\to 0$ is finite and obtain
\begin{equation}\label{eq:alphabeta}
\alpha_{i}^{m} = \beta_{i}^{m} + \left\{\begin{array}{ll} \displaystyle   0 &\qquad\text{($i\not= m+2j$)}\,,\\[0.4cm]\displaystyle 
- c_{j}^{m}\Bigl( \ln 2 + H_{m+2j} + H_{m+j} - 2H_{2m+2j+1} \Bigr) &\qquad\text{($i = m+2j$)}\,,\end{array}\right.
\end{equation}
where the new coefficient $\beta_{i}^{m}$ reads explicitly
\begin{align}\label{eq:betami}
\!\!\beta_{i}^{m} &= \sum_{k=0}^{m-1} {\genfrac{(}{)}{0pt}{}{i}{k}}\,2^k(k!)^2\frac{(m-k-1)!}{(m+k+1)!}
+ \sum_{k=m}^{i} {\genfrac{(}{)}{0pt}{}{i}{k}}\,2^k(k!)^2(-)^{m+k}\frac{H_{k-m} - H_{k+m+1}}{(k-m)!(k+m+1)!}\,,
\end{align}
with the convention that $\genfrac{(}{)}{0pt}{}{i}{k}=0$ whenever $i<k$. Hence our final result for the near zone expansion of $\mathcal{F}_m$ reads 
\begin{align}\label{eq:resFmfinal_app}
& \mathcal{F}_m \stackrel{r/c\to 0}{~=~} \,\sum_{i=0}^{+\infty} \beta_{i}^m\,\frac{(-)^i}{i!}\left(\frac{r}{c}\right)^i F^{(i)}(t)\\
&\qquad + \sum_{j=0}^{+\infty} c_{j}^m\,\frac{(-)^m}{(m+2j)!}\left(\frac{r}{c}\right)^{m+2j}\int_0^{+\infty}\!\!\dd\tau\biggl[\ln\left(\frac{c\tau}{2r}\right) - H_{m+j} + 2 H_{2m+2j+1} \biggr]\, F^{(m+2j+1)}\left(t-\tau\right)\,.\nonumber
\end{align}

\section{Verification of the result~\eqref{eq:resMij} via an alternative procedure}
\label{app:method_2}

The general method exposed in Sec.~\ref{sec:method} suggests a simplifying modification of the MPM algorithm to compute the contributions to the metric of interactions involving a least one gauge moment. In this alternative approach, the construction of the generic MPM metric $h^{\mu\nu}_{\text{gen}}$, of the coordinate shift $\varphi^\mu$ and of the functional relation~\eqref{eq:ML_SL}, between the canonical and the source/gauge moments, is achieved using a variant of the recurrence procedure exposed in Sec.~\ref{sec:implnPM}.

The initial step remains unchanged. Let us now make the recurrence hypothesis at order $n$. Namely, we assume that we have already determined $h^{\mu\nu}_{\text{gen}\, m}$, $\varphi_{m}^\mu$,  $\dM_{m\,L}$, and $\dS_{m\,L}$, for all $m \leqslant n-1$. The equation~\eqref{eq:sourcen} shows that the generic MPM source of the field equations is a sum of three terms. The first one is the source of the canonical metric $\Lambda^{\mu\nu}_{\text{can}\, n}=\Lambda^{\mu\nu}_n [h_{\text{can}\, 1}, \cdots, h_{\text{can}\, n-1}]$. 
%As the latter function is at least quadratic in its arguments, it cannot contain any interaction of order $n$ involving one or more gauge moments. Moreover, 
While this term enters the derivation of $\mathcal{H}^{\mu\nu}_{\text{can}\, n}[\dM_{\leqslant n-1\,L}, \dS_{\leqslant n-1\,L}]$, it does not play any role in the calculation of $\dM_{n\, L}$ and $\dS_{n\, L}$, which are determined by $X^{\mu\nu}_n$ and $Y^{\mu\nu}_n$ hence by $\Delta^{\mu\nu}_n$ and $\Omega^{\mu\nu}_n$. The modification of the recurrence will thus concern the treatment of the second and third terms, which define the reduced source of order $n$:
\begin{align} \label{eq:reduced_source}
\delta \Lambda^{\mu\nu}_{\text{can}\,n} \equiv \Box \Omega^{\mu\nu}_n + \partial \Delta^{\mu\nu}_n \, .
\end{align}

This source is at least quadratic in its arguments $\varphi_1\cdots \varphi_{m-1}, h_{\text{can}\, 1}\cdots h_{\text{can}\, m-1}$, but it is linear in the higher order pieces of the canonical metric or the coordinate shift, $h^{\mu\nu}_{\text{can}\, n-1}$ and $\varphi^\mu_{n-1}$ respectively, which can only enter its quadratic part, through the combination 
\begin{align} \label{eq:quad_source}
\delta \Lambda^{\mu\nu}_{\text{quad}\, n} \equiv \Box\Omega^{\mu\nu}_{1,n-1}+ \Box\Omega^{\mu\nu}_{n-1,1} + \partial \Delta^{\mu\nu}_{1,n-1} + \partial \Delta^{\mu\nu}_{n-1,1} \,,
\end{align}
where the first two terms on the right-hand side belong to $\Omega^{\mu\nu}_n$, and the next two ones belong to $\partial \Delta^{\mu\nu}_n$. Explicitly, they
read
\begin{subequations}
	\begin{align}
	\Omega_{1,n-1}^{\mu\nu} =& - \partial_\rho\left[\varphi_1^\rho\left(h_{\text{can}\, n-1}^{\mu\nu}+\partial\varphi_{n-1}^{\mu\nu}\right)\right]
	+ 2\,\partial_\rho\varphi_1^{(\mu}\,h_{\text{can}\, n-1}^{\nu)\rho}\nonumber\\
	& 
	+ \partial^\rho\varphi_1^{(\mu}\,\partial_\rho\varphi_{n-1}^{\nu)}
	+\frac{1}{2}\eta^{\mu\nu}\left[\partial_\rho\varphi_1^\sigma\partial_\sigma\varphi_{n-1}^\rho-\partial_\rho\varphi_1^\rho\partial_\sigma\varphi_{n-1}^\sigma\right]\,,\\
	%%%%%%%%%%%%%%%%%%%%%%%%%%%%%%%%%%%%%%%%%%%%%%%%%%%%%%%%%%%%%%%
	\Omega_{n-1,1}^{\mu\nu} =& - \partial_\rho\left[\varphi_{n-1}^\rho\left(h_{\text{can}\,1}^{\mu\nu}+\partial\varphi_1^{\mu\nu}\right)\right]
	+ 2\,\partial_\rho\varphi_{n-1}^{(\mu}\,h_{\text{can}\,1}^{\nu)\rho}\nonumber \\
	& 
	+ \partial^\rho\varphi_{n-1}^{(\mu}\,\partial_\rho\varphi_1^{\nu)}	+\frac{1}{2}\eta^{\mu\nu}\left[\partial_\rho\varphi_{n-1}^\sigma\partial_\sigma\varphi_1^\rho-\partial_\rho\varphi_{n-1}^\rho\partial_\sigma\varphi_1^\sigma\right]\,,\\[0.2cm]
	%%%%%%%%%%%%%%%%%%%%%%%%%%%%%%%%%%%%%%%%%%%%%%%%%%%%%%%%%%%%%%%
	\Delta_{1,n-1}^{\mu} =& - h_{\text{can}\,n-1}^{\rho\sigma}\,\partial_{\rho\sigma}\varphi_1^\mu \,,\\[0.2cm]
	%%%%%%%%%%%%%%%%%%%%%%%%%%%%%%%%%%%%%%%%%%%%%%%%%%%%%%%%%%%%%%%
	\Delta_{n-1,1}^{\mu} =& - h_{\text{can}\,1}^{\rho\sigma}\,\partial_{\rho\sigma}\varphi_{n-1}^\mu + \partial_\rho\left(\varphi_1^\rho \Delta_{n-1}^\mu\right)\,.
	\end{align}
\end{subequations}
These relations are straightforward generalizations of Eqs.~\eqref{eq:Omega3}--\eqref{eq:Delta3}. 

The action of the operator $\mathrm{FP}\, \Box^{-1}_\mathrm{ret}$ to $\delta \Lambda^{\mu\nu}_{\text{quad}\, n}$ generates difficult integrals characterized by the presence of $h^{\mu\nu}_{\text{can}\, n-1}$ or $\varphi^\mu_{n-1}$ in the source. The aim of the method presented here is to bypass their evaluation. As part of our recursive hypothesis, we assume that the easier integration of source terms involving for instance $\Omega^{\mu\nu}_{2,n-2}$, $\Omega^{\mu\nu}_{3,n-3}$ \textit{etc.}, has been solved in previous recurrence steps.

Noticing that the partial source term $\delta \Lambda^{\mu\nu}_{\text{quad}\, n}$ is divergenceless, we can apply to it the same treatment as for the right-hand side of Eq.~\eqref{eq:sourcen}, described in Sec.~\ref{sec:implnPM}, with the substitutions $h^{\mu\nu}_{\text{gen} \, n}\to \delta h^{\mu\nu}_{\text{quad}\, n}$, $\Lambda^{\mu\nu}_{\text{can}\, n}\to 0$, $\partial \Delta^{\mu\nu}_{n}+\Box \Omega^{\mu\nu}_{n} \to \delta \Lambda^{\mu\nu}_{\text{quad}\, n}$. This yields
\begin{align}
\delta h^{\mu\nu}_{\text{quad}\, n} = \partial \delta \phi^{\mu\nu}_{\text{quad}\, n} + \Omega^{\mu\nu}_{1,n-1} + \Omega^{\mu\nu}_{n-1,1} + \delta U^{\mu\nu}_{\text{quad}\, n} + \delta V_{\text{quad}\, n}^{\mu\nu}\, , 
\end{align}
with the notations $\delta U^{\mu\nu}_{\text{quad}\, n}=\delta X^{\mu\nu}_{\text{quad}\, n}+\delta Y^{\mu\nu}_{\text{quad}\, n}$ and
\begin{subequations}
	\begin{align}
	\delta \phi_{\text{quad}\, n}^{\mu} &= \FPprop \biggl[
	\left(\frac{r}{r_0}\right)^B (\Delta_{1,n-1}^{\mu}+\Delta_{n-1,1}^{\mu}) \biggr]\,,\\
	\delta X^{\mu\nu}_{\text{quad}\, n} &= \FPprop \biggl[ B \left(\frac{r}{r_0}\right)^B
	\biggl(-\frac{B+1}{r^{2}}\,\bigl(\Omega_{1,n-1}^{\mu\nu}+ \Omega_{n-1,1}^{\mu\nu}\bigr) -
	\frac{2}{r}\,\partial_r \bigl(\Omega_{1,n-1}^{\mu\nu}+ \Omega_{n-1,1}^{\mu\nu}\bigr)\biggr)\biggr] \,,\\
	\delta Y^{\mu\nu}_{\text{quad}\, n} &= 
	\FPprop \biggl[ B \left(\frac{r}{r_0}\right)^B \frac{n_i}{r}
	\biggl( - 2\delta^{i(\mu} \bigl(\Delta_{1,n-1}^{\nu)}+ \Delta_{n-1,1}^{\nu)}\bigr)   +
	\eta^{\mu\nu}\bigl(\Delta_{1,n-1}^i + \Delta_{n-1,1}^i\bigr)\biggr) \biggr]\,;
	\end{align}
\end{subequations}
the term $\delta V^{\mu\nu}_{\text{quad}\, n}$ is computed from the harmonicity algorithm $\delta V_{\text{quad}\, n}^{\mu\nu} = \mathcal{V}^{\mu\nu}[\delta W_{\text{quad}\, n}]$, with $\delta W^{\mu}_{\text{quad}\, n} \equiv \partial_\nu \delta U^{\mu\nu}_{\text{quad}\, n}$. The source of the coordinate shift $\delta \phi^{\mu}_{\text{quad}\, n}$, which involves the $n-1$ order piece of the metric $h^{\mu\nu}_{\text{can}\, n-1}$, is of the same undesirable type as $\delta \Lambda^{\mu\nu}_{\text{quad}\, n}$. However, in the current procedure, the canonical moments will be read off from the gravitational waveform, which is not sensitive to linear-looking gauge transformations. On the other hand, although the commutator contributions do contain terms proportional to $h^{\mu\nu}_{\text{can}\, n-1}$ or $\varphi_{n-1}^\mu$, their brute force integration is not required. Thus, it will be possible to determine $\dM_{L\, n}$ and $\dS_{L\, n}$ without integrating any such term. Once $\delta h^{\mu\nu}_{\text{quad}\, n}$ is obtained, we compute the rest of the metric $h^{\mu\nu}_{\text{rest}\, n} \equiv h^{\mu\nu}_{\text{gen}\, n} - \delta h^{\mu\nu}_{\text{quad}\, n}$ by solving the equation
\begin{align}
\Box h^{\mu\nu}_{\text{rest}\, n} = \delta \Lambda^{\mu\nu}_{\text{rest}\, n} \equiv \Lambda^{\mu\nu}_{\text{can}\, n} - \delta \Lambda^{\mu\nu}_{\text{quad}\, n}\,,
\end{align}
with the help of the standard MPM algorithm. By construction, the source term $\delta \Lambda^{\mu\nu}_{\text{rest}\, n}$ does not depend on $h^{\mu\nu}_{\text{can}\, n-1}$ nor $\varphi_{n-1}^\mu$ and is thus free of the difficult contributions we wanted to avoid. At this stage, the $n$-th order waveform may be built from %$h^{\mu\nu}_{\text{gen}\, \leqslant n} = (h^{\mu\nu}_{\text{gen}\, 1} +\cdots + h^{\mu\nu}_{\text{gen}\, n-1}) + h^{\mu\nu}_{\text{gen}\, n}$, 
$h^{\mu\nu}_{\text{gen}\, \leqslant n} = G h^{\mu\nu}_{\text{gen}\, 1} + \cdots + G^n h^{\mu\nu}_{\text{gen}\, n}$, 
by taking the limit $R\to +\infty$ for constant asymptotically null time $U=T-R/c$, which corresponds to ``radiative'' coordinates $(T=t- 2 G \dM/c^3 \ln (r/r_0), R=r)$~\cite{BlanchetLR}. In those coordinates, the leading order contribution to the metric, $H^{\mu\nu}_{\text{gen}\, \leqslant n}$, admits the same expression as the one in harmonic gauge, but with the logarithms $\ln r$ effectively replaced by $\ln r_0$, and with $(t,r)$ replaced by their radiative counterparts $(T,R)$; see, \textit{e.g.}, Ref.~\cite{FBI15}. This yields in particular [up to terms $\mathcal{O}(R^{-2})$]
\begin{align}
H^{\mu\nu}_{\text{gen}\, n} &=  \Bigl(h^{\mu\nu}_{\text{rest}\, n} + \partial \delta \phi^{\mu\nu}_{\text{quad}\, n}\Bigr)\Big|_{\ln r \to \ln r_0} + \Omega^{\mu\nu}_{1,n-1} + \Omega^{\mu\nu}_{n-1,1} + \delta U^{\mu\nu}_{\text{quad}\, n} + \delta V_{\text{quad}\, n}^{\mu\nu} \,,
\end{align}
where $\partial \delta \phi^{\mu\nu}_{\text{quad}\, n}$ is the linear gauge transformation associated with $\delta\phi^\mu_{\text{quad}\,n}$. The $n$-th order waveform $h^{\text{TT}}_{\leqslant n\, ij}$ is then the transverse trace-free projection of the $1/R$ term in $H^{\mu\nu}_{\text{gen}\, \leqslant n}$. As the TT projection of linear gauge transformations vanishes at order $1/R$, the vector $\delta\phi^\mu_{\text{quad}\,n}$ is actually not required. The result for $h^{\text{rad}}_{\leqslant n\, ij}$ is a certain functional of the source/gauge moments:
\begin{subequations}
\begin{align}\label{eq:hradgen}
h^{\text{rad}}_{\leqslant n\, ij} = \mathcal{H}^{\text{rad}}_{\leqslant n\, ij}[\dI_L, \dJ_L, \cdots, \dZ_L]\, ,
\end{align}
and for the canonical moments, we must also have, at the same time:
\begin{align} \label{eq:hradcan}
h^{\text{rad}}_{\leqslant n\, ij} = \mathcal{H}^{\text{rad}}_{\leqslant n\, ij}[\dM_L, \dS_L, 0, 0, 0, 0]\, .
\end{align}
\end{subequations}

The expressions of $\dM_{n\, L}$ and $\dS_{n\, L}$ are finally found by guess work. We assume they are sums of terms involving the source/gauge moments, %compatible with the symmetries of the problem (translation, rotation and parity), and 
with consistent index structures and physical dimensions, but arbitrary coefficients. Those are fixed by identifying Eq.~\eqref{eq:hradgen} with the outcome that ensues from inserting our ansatz into Eq.~\eqref{eq:hradcan}. Of course, if we wish to iterate the process to the next order $n+1$, we will eventually need to tackle the difficult integrals of source terms containing $h^{\mu\nu}_{\text{can}\, n-1}$, which arise in the calculation of $h^{\mu\nu}_{\text{can}\, n}$ and $\delta \phi^{\mu}_{\text{quad}\, n}$. Nonetheless, we have managed to push this step to the very end.

This strategy is particularly relevant to determine the canonical moments at cubic order for two reasons: (i) The latter task does not demand computing $h^{\mu\nu}_{\text{can}\, 3}$ nor $\delta \phi^\mu_3$, which means that all retarded integrals we have to consider are sourced by functions of $h^{\mu\nu}_{\text{can} \, 1}$, $\varphi^\mu_{1}$, or their derivatives; (ii) The corresponding integrands have the form $f(t-r/c) \hat{n}_L/r^k$, with $k\in \mathbb{N}\setminus\{0,1\}$, whose finite part retarded integral are explicitly known~\cite{B98quad}. In practice, we build the cubic source $\Lambda^{\mu\nu}_{3}$, subtract the ``difficult'' part $\delta \Lambda^{\mu\nu}_{3}$, and apply the MPM algorithm to the rest, which leads to $h^{\mu\nu}_{\text{rest}}$. At last, we compute the commutators $\delta U^{\mu\nu}_{\text{quad}\, 3}$ with the method developed in Sec.~\ref{sec:retarded_integrals}, from which we can infer $\delta V^{\mu\nu}_{\text{quad}\, 3}$. The cubic waveform follows from the effective metric
\begin{align}
h^{\mu\nu}_{\text{eff}\, \leqslant 3} = G h^{\mu\nu}_{\text{gen}\, 1} + G^2 h^{\mu\nu}_{\text{gen}\, 2}  + G^3 \left[ h^{\mu\nu}_{\text{rest}\, 3} + \Omega^{\mu\nu}_{12} + \Omega^{\mu\nu}_{21} + \delta U^{\mu\nu}_3\right]\,.
\end{align}
Our final result~\eqref{eq:resMij} for the canonical quadrupole moment was obtained following the general method in Secs.~\ref{sec:general_method}--\ref{sec:implnPM}, and has then been entirely checked using this approach.

\bibliography{ListeRef_jauge.bib}

\end{document}